\documentclass[eqsecnum,twocolumn,tightenlines,floats,floatfix,prc,nofootinbib,preprintnumbers]{revtex4}%

\def\bea{\begin{eqnarray}}
\def\eea{\end{eqnarray}}

\def\st#1{{\kern-4pt} \not\!#1}

\def\sp{\kern +3pt}
\def\sm{\kern -3pt}

\def\be{\begin{equation}}
\def\ee{\end{equation}}
\def\ba{\begin{eqnarray}}
\def\ea{\end{eqnarray}}

\usepackage{graphics}
\usepackage{graphicx}
\usepackage{epsf} 
\usepackage{amsmath}
\usepackage{amssymb}
\usepackage{slashed}
\usepackage{bm}

\def\sfrac#1#2{{\textstyle \frac{#1}{#2}}}

\begin{document}

\phantom{0}
\vspace{-0.2in}
\hspace{5.5in}

% include preprint number option
%\preprint{2014.XXX}
\preprint{}

\vspace{-1in}%\parbox{1.5in}{ \vspace{-9.6in}}  % moves the preprint box down

\title
{\bf 
Using the single quark transition model to 
predict nucleon \\ resonance amplitudes}
\author{G.~Ramalho 
\vspace{-0.1in}  }

\affiliation{
International Institute of Physics, Federal 
University of Rio Grande do Norte, Avenida Odilon Gomes de Lima 1722, 
Capim Macio, Natal-RN 59078-400, Brazil
%\vspace{-0.15in}
}

\vspace{0.2in}
\date{\today}

\phantom{0}

\begin{abstract}
We present predictions for the 
$\gamma^\ast N \to N^\ast$ helicity amplitudes,
where $N^\ast$ is a member of the
$[70,1^-]$ supermultiplet. 
We combine the results from the 
single quark transition model for the helicity 
amplitudes with the results of the covariant spectator quark 
model for the $\gamma^\ast N \to N^\ast(1535)$ 
and $\gamma^\ast N \to N^\ast(1520)$ transitions.
The theoretical estimations from the  covariant spectator quark model
are used to calculate three independent functions $A,B$, and $C$
of $Q^2$, where 
$Q^2=-q^2$ and $q$ is the momentum transfer. 
With the knowledge of the functions $A,B$, and $C$
we estimate the helicity amplitudes for the 
transitions
$\gamma^\ast N \to N^\ast(1650)$,  
$\gamma^\ast N \to N^\ast(1700)$,
$\gamma^\ast N \to \Delta(1620)$,
and $\gamma^\ast N \to \Delta(1700)$.
The analysis is restricted to reactions with proton targets.
The predictions for the transition amplitudes
are valid for $Q^2 > 2$ GeV$^2$.
\end{abstract}

%\phantom{0}
%\vspace{7.0in}
%\vspace{-6in}
\vspace*{0.9in}  % sets how far the title is below the preprint box
\maketitle

\section{Introduction}

One of the challenges of modern physics 
is the description of the internal structure of the hadrons.
It is believed that the substructure of the the hadrons 
in general and the nucleon and the nucleon resonances
in particular,  are ruled by Quantum Chromodynamics (QCD), 
in terms of quark and gluon degrees of freedom.
Although QCD can be useful for reactions at 
high  $Q^2$, 
it becomes more complex at 
low and intermediate $Q^2$,
which restrains 
the theoretical predictions for that range~\cite{Aznauryan12,NSTAR}.
Therefore in practice to obtain 
predictions for small $Q^2$ one has sometimes
to rely on effective degrees of freedom  
as the constituent quarks.

The quark substructure of a baryon can in first approximation 
be classified in terms of the $SU(6)$ spin-flavor symmetry,
combined with the $O(3)$ group for radial and 
rotational excitations.
In that framework
the spin 1/2 baryons, including the nucleon, 
and the spin 3/2 baryons can be classified 
in supermultiplets $[SU(6),L^P]$ characterized by 
angular momentum ($J$), quark total spin 
($S=1/2,3/2$),
orbital angular momentum ($L$) and parity ($P$).
In the notation $[SU(6),L^P]$, $SU(6)$ represents the number 
of particles of the multiplet (including all spin projections). 
Then the nucleon ($J^P=\sfrac{1}{2}^+$) is part of the $[56,0^+]$ 
supermultiplet
and the states $N^\ast(1535)$ ($J^P=\sfrac{1}{2}^-$), 
also represented by $S_{11}(1535)$, and 
$N^\ast(1520)$ ($J^P=\sfrac{3}{2}^-$), 
also represented by $D_{13}(1520)$ are part of 
the $[70,1^-]$ supermultiplet \cite{Capstick00}.

The use of the $SU(6)\otimes O(3)$ group \cite{Koniuk80,Capstick00} 
to represent 
the wave functions of a baryon (three-quark system)
combined with 
the electromagnetic interaction in impulse approximation 
leads to the so-called single quark transition model (SQTM) 
\cite{Hey74,Cottingham79,Burkert03}.
Here, single means 
that only one quark couples with the photon (impulse approximation).
In these conditions the SQTM can be used to 
parametrize the transition current 
between two supermultiplets,
in an operational form 
that includes only four independent terms,
with coefficients exclusively dependent of $Q^2$.

In particular, the SQTM can be used to parametrize 
the $\gamma^\ast N \to N^\ast$  transitions,
where $N^\ast$ is a nucleon (isospin 1/2) or a $\Delta$ 
(isospin 3/2) excitation from the
$[70,1^-]$ supermultiplet, in terms 
of three independent functions of $Q^2$:
$A,B$, and $C$ \cite{Hey74,Cottingham79,Burkert03,Burkert04,Aznauryan12}.
The relation between the functions $A,B$, and $C$
and the amplitudes are presented in Table~\ref{tableAmp}.
In the table, besides the transitions $\gamma^\ast N \to S_{11}(1535)$
and $\gamma^\ast N \to D_{13}(1520)$,
one has expressions  for the transitions
$\gamma^\ast N \to S_{11}(1650)$, 
$\gamma^\ast N \to D_{13}(1700)$,
$\gamma^\ast N \to S_{31}(1620)$, and 
$\gamma^\ast N \to D_{33}(1700)$.
Once the coefficients $A,B$, and $C$
are determined it is possible to predict 
the transition helicity amplitudes for all 
the resonances from the $[70,1^-]$ supermultiplet.
The relations presented in the table are based in the 
exact $SU(6)$ spin-flavor symmetry
broken by the color-hyperfine interaction between quarks.
That interaction leads to the configuration 
mixing between various baryon states characterized 
by some mixing angles,
estimated from hadron decays~\cite{Aznauryan12,Capstick00,Koniuk80,Burkert03}.
Note however, that as the SQTM is based exclusively 
on the valence quark degrees of freedom, 
we should not expect a good description of the 
reactions at low $Q^2$, where meson cloud effects may be very 
important~\cite{Aznauryan12,NSTAR,Burkert04,Tiator04,Chen07,Capstick07,Dong14,OctetFF,Octet2Decuplet}.
Calculations for the same helicity amplitudes 
using quark models can be found in 
Refs.~\cite{Aznauryan12,Burkert04,Tiator04,Warns90,Capstick95,Bijker96,Aiello,Bijker98,Santopinto98,Pace00,Merten02,Aznauryan12b,Santopinto12,Ronniger13,Golli}.

The covariant spectator quark model 
was applied in the past 
to the electromagnetic structure 
of the nucleon \cite{Nucleon} 
and the $S_{11}(1535)$, $D_{13}(1520)$ excitations~\cite{S11,D13}.
The determination of the transition helicity amplitudes 
are based  mainly on the valence quark content, 
but some information about additional effects like the 
meson cloud dressing can be inferred from the formalism.
One can then use the results 
from the covariant spectator quark model
for the $A_{1/2}$ amplitude 
in the  $S_{11}(1535)$ transition and the two transverse amplitudes 
($A_{1/2}$ and $A_{3/2}$) in the $D_{13}(1520)$ transition,
to calculate $A,B$, and $C$.
Note, however, that because the valence quark effects 
are dominant only at large $Q^2$, 
the results are accurate only in that region.
Based on the results of the 
covariant spectator quark model for the 
$\gamma^\ast N \to S_{11}(1535)$ 
and $\gamma^\ast N \to D_{13}(1520)$ transitions
we estimate that the 
predictions of the model should be 
accurate for the $Q^2 > 2$ GeV$^2$ region.

As the covariant spectator quark model breaks 
the $SU(2)$-isospin symmetry,
the use of that model to calculate the 
functions $A,B$, and $C$ from the SQTM 
has to be understood as 
an approximation, with a degree of error proportional 
to the  percentage of  the  $SU(2)$ breaking.
As consequence the estimation 
for neutral reactions (with neutron targets) 
will be less reliable, since in the covariant
spectator quark model those reactions 
depend significantly
of the $SU(2)$ breaking.
For instance, the neutron electric form 
factor would vanish if the  
$SU(2)$ symmetry breaking was not considered \cite{Nucleon}.

The article is organized as follows:
In the next section we present the relations 
between the $\gamma^\ast N \to N^\ast$ amplitudes 
for the  $N^\ast$ resonances of the $[70,1^-]$ supermultiplet 
and the functions $A,B$, and $C$, according to the SQTM.
In Sec.~\ref{secCSQM} we discuss the 
formalism of the covariant spectator quark model.
The expressions of the  covariant spectator quark model
for the  $S_{11}(1535)$ and  $D_{13}(1520)$ excitations 
are presented in Sec.~\ref{secAmps}.
The numerical results for the $[70,1^-]$ amplitudes 
are presented in Sec.~\ref{secResults}.
The summary and the conclusions are in Sec.~\ref{secConclusions}.

\section{Determination of the functions  $A,B$, and $C$}
\label{secSQTM}

The expressions for the $\gamma^\ast N \to N^\ast$ amplitudes
for a $N^\ast$ resonance from the $[70,1^-]$ 
supermultiplet calculated by the SQTM \cite{Burkert03,Burkert04}
are presented in Table~\ref{tableAmp}.
For the mixing angles we use 
the values from Ref.~\cite{Burkert03}:
$\theta_S=  31^\circ$ and $\theta_D= 6^\circ$. 
Since the SQTM estimates are derived 
from the interaction of quarks 
with transverse photons there are 
no estimates for the amplitudes $S_{1/2}$~ \cite{Burkert03}.
Using the table, we can write 
in particular for the $S_{11}(1535)$ (label $S11$)
and $D_{13} (1520)$ (label $D13$) cases:
\ba
A_{1/2}^{S11}= \frac{1}{6}(A+B-C) \cos  \theta_S,
\ea
and 
\ba
& &
A_{1/2}^{D13}=
\frac{1}{6\sqrt{2}}(A-2B-C) \\
& &
A_{3/2}^{D13}= 
\frac{1}{2\sqrt{6}}(A+ C),
\label{eqA32_D13}
\ea
where in the last expressions we 
approximate $\cos \theta_D =0.99 \to 1$.

From the previous relations, we obtain
\ba
& &
A= 2 \frac{A_{1/2}^{S11}}{\cos \theta_S} 
+ \sqrt{2} A_{1/2}^{D13} +
\sqrt{6} A_{3/2}^{D13} 
\label{eqA}
\\
& & 
B=   2 \frac{A_{1/2}^{S11}}{\cos \theta_S} 
- 2 \sqrt{2} A_{1/2}^{D13} 
\label{eqB}
\\
& &
C= - 2 \frac{A_{1/2}^{S11}}{\cos \theta_S} 
- \sqrt{2} A_{1/2}^{D13} +
\sqrt{6} A_{3/2}^{D13}. 
\label{eqC}
\ea

An interesting approximation is the case $A_{3/2}^{D13} \simeq 0$.
From Eq.~(\ref{eqA32_D13}) we conclude that 
the approximation is equivalent 
to $A+C \simeq 0$, 
or $C \simeq -A$, reducing the number of 
functions to be determined to only 2 ($A$ and $B$).
In the case $A_{3/2}^{D13} \simeq 0$,
we obtain then
\ba
& &A_{1/2}^{S11} \simeq \frac{1}{6}(2A+B) \cos  \theta_S, 
\label{eqA12S1}
\\
& &A_{1/2}^{D13} \simeq
\frac{\sqrt{2}}{6}(A-B). %\\& &A_{3/2}^{D13} \simeq 0.
\label{eqA12D1}
\ea

\begin{table}[t]
\begin{tabular}{c c c }
\hline
\hline
State & Amplitude &     \\
\hline
$S_{11}(1535)$ & $A_{1/2}$ & $\frac{1}{6}(A+B-C) \cos  \theta_S$ \\ [.3cm]
% & & \\
$D_{13}(1520)$
& $A_{1/2}$ & $\frac{1}{6\sqrt{2}}(A-2B-C)\cos \theta_D$   \\
& $A_{3/2}$ & $\frac{1}{2\sqrt{6}}(A+ C) \cos \theta_D$   \\ [.3cm]
% & & \\
$S_{11}(1650)$ & $A_{1/2}$ & $\frac{1}{6}(A+B-C) \sin  \theta_S$ \\ [.3cm]
$S_{31}(1620)$ & $A_{1/2}$ & $\frac{1}{18}(3A-B+C) $ \\ [.3cm]
$D_{13}(1700)$ & $A_{1/2}$ &  $\frac{1}{6\sqrt{2}}(A-2B-C)\sin \theta_D$  \\
& $A_{3/2}$ & $\frac{1}{2\sqrt{6}}(A+ C) \sin \theta_D$   \\ [.3cm]
$D_{33}(1700)$ & $A_{1/2}$ &  
$\frac{1}{18 \sqrt{2}}(3A+2B+C)$  \\ 
& $A_{3/2}$ & $\frac{1}{6\sqrt{6}}(3A-C) $   \\ [.1cm]
\hline
\hline
\end{tabular}
\caption{
Amplitudes $A_{1/2}$ and $A_{3/2}$ estimated by SQTM 
for the proton targets ($N=p$).
The angle $\theta_S$ is the mixing angle associated 
with the $S_{11}$ states ($\theta_S =  31^\circ$).
The angle $\theta_D$ is the mixing angle associated 
with the $D_{13}$ states ($\theta_S = 6^\circ$).}
\label{tableAmp}
\end{table}

The study of the 
$\gamma^\ast N \to D_{13}(1520)$ transition suggests that 
the amplitude $A_{3/2}$ falls off  
faster than $A_{1/2}$ with $Q^2$,  
justifying the approximation $A_{3/2} \simeq 0$
for large $Q^2$~\cite{D13}.
In the covariant spectator quark model,
in particular  $A_{3/2}^{D13} \approx 0$ 
when the meson cloud effects
are not included.
Therefore, in
that model the results for $A_{3/2}^{D13}$
are  interpreted as the exclusive consequence 
of the meson cloud effects. 
However, the falloff of $A_{3/2}$ 
is slow when compared with the typical 
falloff from the meson cloud effects~\cite{D13}.
In order to check if our estimate 
can be improved in this paper we include also 
a parametrization for the  amplitude $A_{3/2}^{D13}$,
which simulates the meson cloud effects.

\section{Covariant spectator quark model}
\label{secCSQM}

In the covariant spectator quark model,
baryons are treated as
three-quark systems. 
The baryon wave functions are derived from 
the quark states according to the  
$SU(6) \otimes O(3)$ symmetry group.
A quark is off-mass-shell, and  free
to interact with the photon fields,
and other two quarks are on-mass-shell~\cite{Nucleon,Nucleon2,OctetFF,Omega}.
Integrating over the quark-pair degrees
of freedom we reduce the baryon to a quark-diquark system,
where the diquark can be represented as
an on-mass-shell spectator particle with an effective
mass of $m_D$~\cite{Nucleon,Nucleon2,Omega,D13}.

The electromagnetic interaction with the baryons
is described by the photon coupling with
the constituent quarks in the relativistic impulse approximation,
and the quark electromagnetic structure is
represented in terms of the quark form
factors parametrized by a vector meson dominance
mechanism~\cite{Nucleon,Omega,Lattice}.
The parametrization of the quark current
was calibrated in the studies of the nucleon form factors~\cite{Nucleon},
by the lattice QCD data for the decuplet baryons~\cite{Omega},
and encodes effectively the gluon
and quark-antiquark substructure of the constituent quarks.

The quark current has the general form~\cite{Nucleon,Omega}
\ba
j_q^\mu(Q^2) = j_1(Q^2) \gamma^\mu
+ j_2(Q^2) \frac{i \sigma^{\mu \nu} q_\nu}{2M},
\label{eqJq}
\ea
where $M$ is the nucleon mass and
$j_i$ $(i=1,2)$ are the Dirac and Pauli quark form factors.
In the $SU(2)$-flavor sector
the functions $j_i$ can also be decomposed 
into the isoscalar ($f_{i+}$) and the isovector ($f_{i-}$) 
components  
\ba
j_i (Q^2)= \frac{1}{6} f_{i+}(Q^2) + \frac{1}{2} f_{i-}(Q^2) \, \tau_3,
\ea
where $\tau_3$ acts on the isospin states of baryons
(nucleon or resonance).
The details can be found in Refs.~\cite{Nucleon,OctetFF,Omega}.
Since the quark current includes a Pauli term,
the quarks have nonzero
anomalous magnetic moment ($\kappa_q$)
in the present formalism.

In the study of inelastic reactions 
(the final state has a mass different from 
the initial state) we replace
 $\gamma^\mu \to \gamma^\mu - \frac{\not q q^\mu}{q^2}$
in Eq.~(\ref{eqJq}).
This procedure is equivalent to the use of the Landau 
prescription in the transition current
and ensures the conservation 
of the transition current between the 
baryon states~\cite{Kelly98,Batiz98,Gilman02}.
The term restores current conservation but does not affect
the results of the observables \cite{Kelly98}.

When the nucleon 
($\Psi_N$) and the  final resonance $R$  ($\Psi_R$),
where $R$ stands for a $N^\ast$ nucleon resonance,
wave functions are 
written in terms of
the single quark and quark-pair states,
the transition current 
can be written in the 
relativistic impulse approximation~\cite{Nucleon,Nucleon2,Omega} as
\ba
J^\mu=
3 \sum_{\Gamma} 
\int_k \bar \Psi_R (P_+,k) j_q^\mu \Psi_N(P_-,k),
\label{eqJmu}
\ea  
where $P_-,P_+$, and $k$ are  the
nucleon, the resonance, and the diquark momenta respectively.
In the previous equation 
the index $\Gamma$ labels
the possible states of the intermediate diquark,
the factor 3 
takes account of the contributions from
the other quark pairs by the symmetry, and the integration
symbol represents the covariant integration over the 
diquark on-mass-shell momentum.

In the calculation of the transition current 
it is convenient to project the states 
on the isospin symmetric components (label $S$) 
or the isospin antisymmetric components (label $A$). 
We can define then, the 
$Q^2$ dependent coefficients
\ba
& &j_i^A= \frac{1}{6} f_{1+} + \frac{1}{2} f_{1-} \tau_3 \\
& &j_i^S= \frac{1}{6} f_{1+} - \frac{1}{6} f_{1-} \tau_3. 
\ea
See Refs.~\cite{Nucleon,OctetFF,Omega} for more details.
For future discussion we note that 
\ba
j_i^A + \frac{1}{3} j_i^S = 
\frac{2}{9}(f_{i+} + 2 f_{i-} \tau_3).
\ea

Using Eq.~(\ref{eqJmu}), we can express 
the transition current 
in terms of the coefficients $j_i^{A,S}$
and the radial wave functions 
$\psi_N$ and $\psi_R$~\cite{Nucleon,S11,D13}.
The radial wave functions are scalar functions that 
depend on the  baryon and diquark  momenta.
Those functions parametrize the momentum distributions 
of the quark-diquark systems. 
From the transition current we can extract 
the form factors and the helicity transition amplitudes,
defined in the rest frame of the resonance (final state), 
for the reaction under study~\cite{Aznauryan12,NSTAR,S11,D13}.

As mentioned, the representation of the quark current 
in terms of a vector meson dominance
parametrization~\cite{Nucleon,OctetFF,Omega}
simulates in an effective way 
the internal structure of the constituent quarks,
including the meson cloud dressing of the quarks.
There are however some processes such as
the meson exchanged between the different quarks
inside the baryon, which cannot be reduced
to simple diagrams with quark dressing.  
Those processes are regarded  
as arising from a meson exchanged between
the different quarks inside
the baryon
and can be classified as meson cloud corrections 
to the hadronic reactions~\cite{OctetFF,Octet2Decuplet,D13}.

The covariant spectator quark model was already 
applied to the $\Delta(1232)$ system 
\cite{NDelta,DeltaSystem}, to some nucleon resonances like 
the Roper, $N^\ast(1520), N^\ast(1535)$, $N^\ast(1710)$
and $\Delta(1600)$
\cite{S11,D13,Resonances,N1710} and  
several reactions with strange 
baryons~\cite{Octet2Decuplet,Strange,LambdaS}.

In the present work the necessary input 
from the covariant spectator quark model is 
the transition form factors 
(that can be rewritten as helicity amplitudes)
for the $\gamma^\ast N \to N^\ast$ 
transitions with $N^\ast=S_{11}(1535), D_{13}(1520)$.
The $\gamma^\ast N \to S_{11}(1535)$ and 
$\gamma^\ast N \to D_{13}(1520)$ transitions
were analyzed in  Refs.~\cite{S11,D13}.
In those papers we used the parametrization 
of the nucleon system given by Ref.~\cite{Nucleon},
which requires  two parameters to describe 
the radial wave function $\psi_N$,
and the parametrization of the quark current described below.
In order to obtain a complete representation 
of the systems $S_{11}(1535)$ and  $D_{13}(1520)$,
one has to define convenient 
radial wave functions that ensures 
the orthogonality between those 
wave functions with the nucleon wave function. 
The subject is discussed in the Appendix.
%Appendix \ref{appS11} 
for the $S_{11}(1535)$ case, and in Ref.~\cite{D13} 
for the $D_{13}(1520)$ case. 
In simple words we can say that we 
define the $N^\ast$ radial wave functions 
with the same long range parametrization 
as the nucleon 
and define a new short range parameter 
for each resonance.
Therefore, we add to the model 
of the nucleon  one new parameter for resonance.
For the case of the $D_{13}(1520)$, 
we include a simple parametrization of the 
meson cloud, as discussed in Ref.~\cite{D13},
in order to reproduce the amplitude $A_{3/2}$.

\section{Parametrization of $S_{11}(1535)$
and  $D_{13}(1520)$ amplitudes }
\label{secAmps}

We will discuss now the parametrizations 
of the amplitudes associated with the 
$\gamma^\ast N \to S_{11}(1535)$ 
and $\gamma^\ast N \to D_{13}(1520)$ transitions.
To distinguish between the two cases we 
will use 
the label $S$ (or $S11$)
for the $S_{11}$ state, and the label $D$ (or $D13$)
for the  $D_{13}$ state.
Then $M_S$ represents the $S11$ mass ($\approx 1.535$ GeV)
and  $M_D$ represents the $D13$ mass ($\approx 1.520$ GeV).
The details of the structure of those systems 
can be found in Refs.~\cite{S11,D13}.
Here we will discuss only the main features 
of those transitions.

For the 
$\gamma^\ast N \to S_{11}(1535)$ we consider 
the calculation from Ref.~\cite{S11},
developed for the high $Q^2$ region, 
that we extend in the present work also 
to the low $Q^2$ region.
The details are presented in the Appendix.
%Appendix \ref{appS11}.
Recall that 
the $S_{11}(1535)$ state is described in the present model 
using exclusively the valence quark degrees of freedom.
The interesting properties 
of the $S_{11}(1535)$ amplitudes are also discussed
in Refs.~\cite{S11scaling,LambdaS}.

For the $\gamma^\ast N \to D_{13}(1520)$ transition
we use the model from Ref.~\cite{D13}, 
particularly for the valence quark contributions.
Since one of the amplitudes ($A_{3/2}$) vanishes 
in the covariant spectator quark model formalism,
when only the valence quark contributions 
are taken into account,  
we investigate also the impact 
of considering a meson cloud parametrization 
for that amplitude.

\subsection{Resonance $S_{11}(1535)$}

The electromagnetic structure of the 
$\gamma^\ast N \to S_{11}(1535)$ transition
can be parametrized by two independent form factors 
$F_1^\ast$ and $F_2^\ast$~\cite{S11,S11scaling}.
The experimental data suggests that $F_2^\ast \simeq 0$ 
for large $Q^2$, more specifically for $Q^2 > 1.5$ GeV$^2$.
As for the $F_1^\ast$, the data are well 
described by the covariant spectator quark model.
Combining both results, 
for large $Q^2$ we can calculate $A_{1/2}$ using 
\ba
A_{1/2}&=&-
\frac{\sqrt{2}}{3} 
F_S\left(
f_{1+} + 2 f_{1-} \tau_3 \right) {\cal I}_{S11} \cos \theta_S,
\label{eqA12S}
\ea 
where
\ba
{\cal I}_{S11} (Q^2)=  \int_k  
\frac{k_z}{|{\bf k}|}
\psi_{S11}(P_+,k) \psi_N(P_-,k), 
\label{eqIS11}
\ea
and
\ba
F_S= 2 e \sqrt{\frac{(M_S + M)^2+Q^2}{8M(M_S^2-M^2)}}.
\ea
In the previous equations ${\cal I}_{S11}$ gives 
an integral defined in the $S11$ rest frame, 
but it can also be written in a covariant form \cite{S11}.
In the $S11$ rest frame one has $P_+= (M_S,0,0,0)$ 
and $P_-=(\sqrt{M^2 + |{\bf q}|^2}, 0,0 ,-|{\bf q}|)$,
where $|{\bf q}|$ is the magnitude of the photon three-momentum.

As mentioned already, the experimental result for
$F_2^\ast $ vanishes for $Q^2 > 1.5$ GeV$^2$ \cite{S11}.
This can be interpreted as a consequence of the cancellation 
between valence quark  and meson cloud  effects~\cite{S11,LambdaS,S11scaling}.
Therefore, although the estimate from the 
valence quark contribution is nonzero, 
if we want to estimate the final result for $F_2^\ast$, 
the best approximation is $F_2^\ast =0$.
A consequence of the previous result 
is that we obtain the best estimate for $A_{1/2}$ when 
we  neglect $F_2^\ast$ in the calculations.

At the moment we discuss the $S_{11}(1535)$ state 
under the assumption that the contributions with 
core spin 1/2 (core spin is the sum of the quark spins),
are the more important components of the wave function
(contributions proportional to $\cos \theta_S$).
However, the $S_{11}(1535)$ state has 
also contributions from states with core spin 3/2 
(proportional to $\sin \theta_S$).
Those contributions were not calculated 
at the moment in the context of the 
covariant spectator quark model framework,
although that can be done in the future,
using the formalism developed in Ref.~\cite{D13}.

As we did not include the  possible effects of 
the core spin 3/2 component, 
it may happen that we are underestimating 
the magnitude of the amplitude $A_{1/2}$.
However, the explicit inclusion of those 
effects would also reduce the previous result
from the spin 1/2 component
from Ref.~\cite{S11}, since we need to correct 
that value by $\cos \theta_S$, 
because the limit $\cos \theta_S=1$
was used in that work.
To accommodate the  core spin 3/2 components 
in an effective way we simply take 
Eq.~(\ref{eqA12S}) with $\cos \theta_S =1$.
Future calculations can test the previous assumption.
However, for the propose of the present work 
we did not expect that the results 
would  be significantly affected by the explicit 
inclusion of the terms with $\sin \theta_S$, which are 
omitted in the present calculation.

\subsection{Resonance $D_{13}(1520)$}

The valence quark contributions for the 
$\gamma^\ast N \to D_{13}(1520)$ 
form factors  can be written as \cite{D13}
\ba
\hspace{-.9cm}
& &
G_M = \frac{1}{3\sqrt{3}} \frac{M}{M_D-M}
\sqrt{\frac{(M_D-M)^2 + Q^2}{(M_D+M)^2+ Q^2}} 
\nonumber \\
\hspace{-.9cm}
& & %\qquad 
\times 
\left[
\left(f_{1+} + 2 f_{1-} \tau_3\right)
+ \frac{M_D+M}{2M} 
\left(  f_{2+} + 2 f_{2-} \tau_3 \right)
 \right] {\cal I}_{D13}  , \nonumber \\
\hspace{-.9cm}
& & \label{eqGM} \\
\hspace{-.9cm}
& &
G_E=-G_M,
\ea
where 
\ba
{\cal I}_{D13} (Q^2)= \int_k
\frac{k_z}{|{\bf k}|}
%\frac{\varepsilon_{0P_+} \cdot \tilde k}{\sqrt{- \tilde k^2}}
\psi_{D13}(P_+,k) \psi_N(P_-,k). 
\ea
The expression for the Coulomb quadrupole form factor $G_C$
is not 
relevant for the present work,
since the SQTM expressions do not apply 
to the $S_{1/2}$ amplitude.
Also, in the last case the integral is 
represented in the resonance $D13$ rest frame:
$P_+= (M_D,0,0,0)$ 
and $P_-=(\sqrt{M^2 + |{\bf q}|^2}, 0,0 ,-|{\bf q}|)$,
where $|{\bf q}|$ is the magnitude of the photon  three-momentum.

From the form factors, we can calculate the 
helicity amplitudes, using \cite{D13}
\ba
& &
A_{1/2}= {F_D} \, G_M + \frac{1}{4}F_D \, G_4^\pi 
\label{eqA12}\\
& &
A_{3/2}= \frac{\sqrt{3}}{4} F_D \, G_4^\pi 
\label{eqA32}
\ea
where 
\ba
F_D=  \frac{e}{M} \sqrt{\frac{M_D-M}{M_D+M}} 
\sqrt{\frac{(M_D+M)^2 + Q^2}{2M}}.  
\ea
In the Eqs.~(\ref{eqA12})-(\ref{eqA32}), 
$G_4^\pi$ is a new function, which vanishes
if only the valence quark contributions are considered,
but that can be used to parametrize 
the pion and other meson cloud effects.

The fit to the $A_{3/2}$ data gives
\ba
G_4^\pi=
1.354 
\left(\frac{\Lambda_4^2}{\Lambda_4^2 + Q^2}
\right)^3 F_\rho,
\label{eqG4}
\ea
where 
$\Lambda_4^2=20$ GeV$^2$.
The function $F_\rho$ is defined as
\ba
F_\rho = \frac{m_\rho^2}{m_\rho^2 + Q^2  + 
\frac{1}{\pi} \frac{\Gamma_\rho^0}{m_\pi} Q^2 \log \frac{Q^2}{m_\pi^2}}.
\ea
In the last expression,
$m_\rho$ and $m_\pi$ are the $\rho$ and pion mass, 
respectively and 
$\Gamma_\rho^0= 0.149$ GeV  \cite{D13}.

Note that the parametrization of the amplitude
$A_{3/2}$ given by Eqs.~(\ref{eqA32}) and (\ref{eqG4}) 
is in the region $Q^2=5$--10 GeV$^2$, 
dominated by the function $F_\rho \propto 1/(Q^2 \log Q^2)$
due to the large cutoff ($\Lambda_4^2=20$ GeV$^2$) 
in the tripole factor.
Therefore in the intermediate $Q^2$ region, 
$A_{3/2}$ shows a slow falloff, 
contrary to what we would expect from 
a meson cloud contribution.

It is worth mentioning that other quark models 
predict nonzero contributions for 
the amplitude $A_{3/2}$ \cite{Warns90,Capstick95,Bijker96,Aiello,Bijker98,Santopinto98,Merten02,Santopinto12,Ronniger13,Golli}.
However, those estimates are small in general, 
and about  20-40\% of the measured 
values~\cite{Warns90,Aiello,Bijker98,Santopinto98,Merten02,Santopinto12,Ronniger13,Golli}.
Those results can be interpreted as an indication 
that the meson cloud effects are the dominant 
contribution for the amplitude $A_{3/2}$,
as assumed in Ref.~\cite{D13} in the 
context of the covariant spectator quark model.
Also the results of the 
Excited Baryon Analysis Center
(EBAC) coupled-channel reaction model 
supports the idea that the meson cloud 
effects are the dominant contribution for the 
amplitude $A_{3/2}$~\cite{EBAC}.

\begin{figure}[t]
\vspace{.3cm}
\centerline{
\mbox{
\includegraphics[width=2.8in]{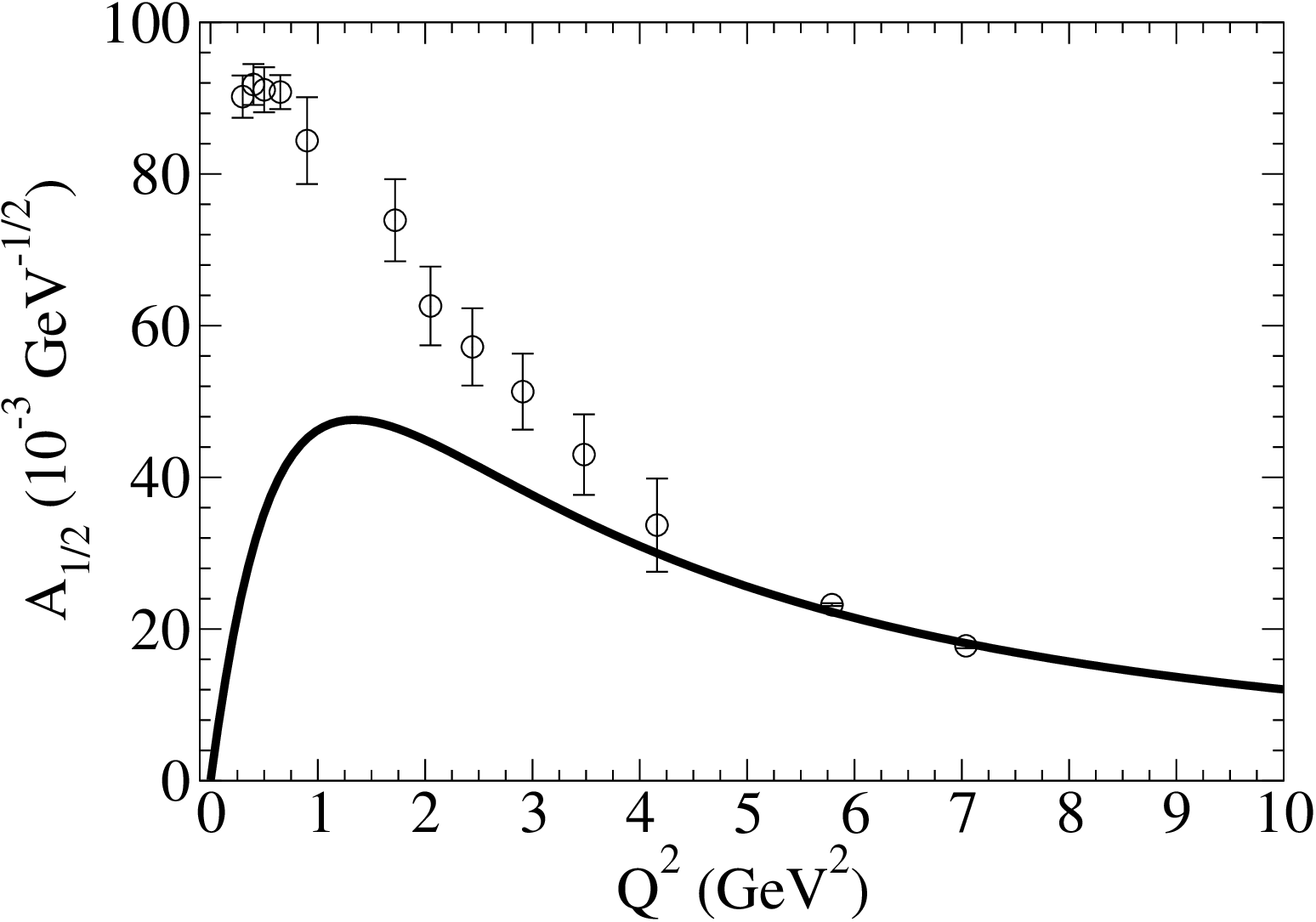}
}}
\caption{\footnotesize{Results for the 
$\gamma^\ast p \to S_{11}^+(1535)$ amplitude $A_{1/2}$.
Datapoints from CLAS~\cite{Aznauryan09}
and JLab/Hall C~\cite{Dalton09}.
}}
\label{figParam1}
\end{figure}
\begin{figure}[t]
\vspace{.3cm}
\centerline{
\mbox{
\includegraphics[width=2.8in]{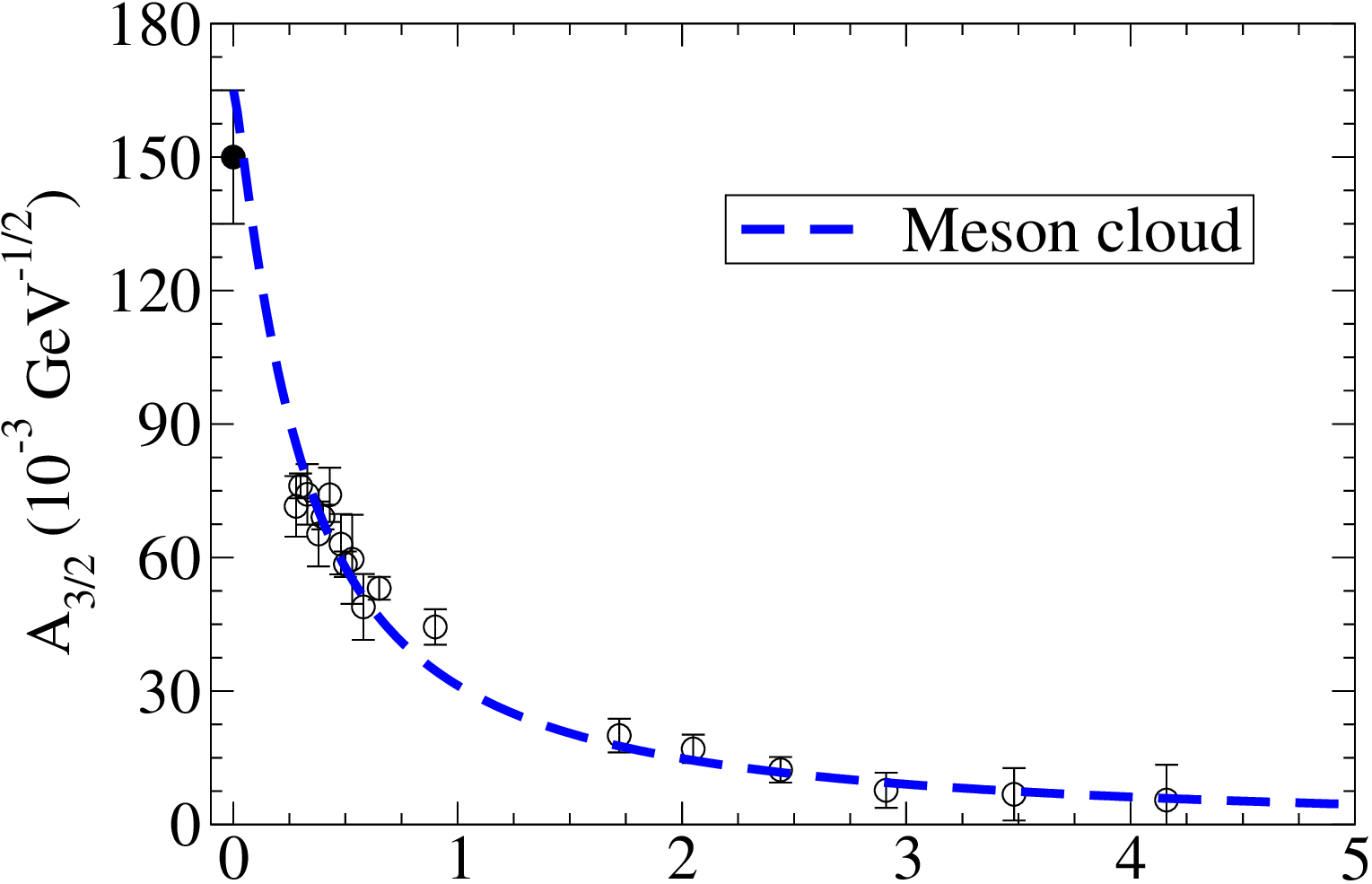}
}}
\centerline{
\vspace{.5cm} }
\centerline{
\mbox{
\includegraphics[width=2.8in]{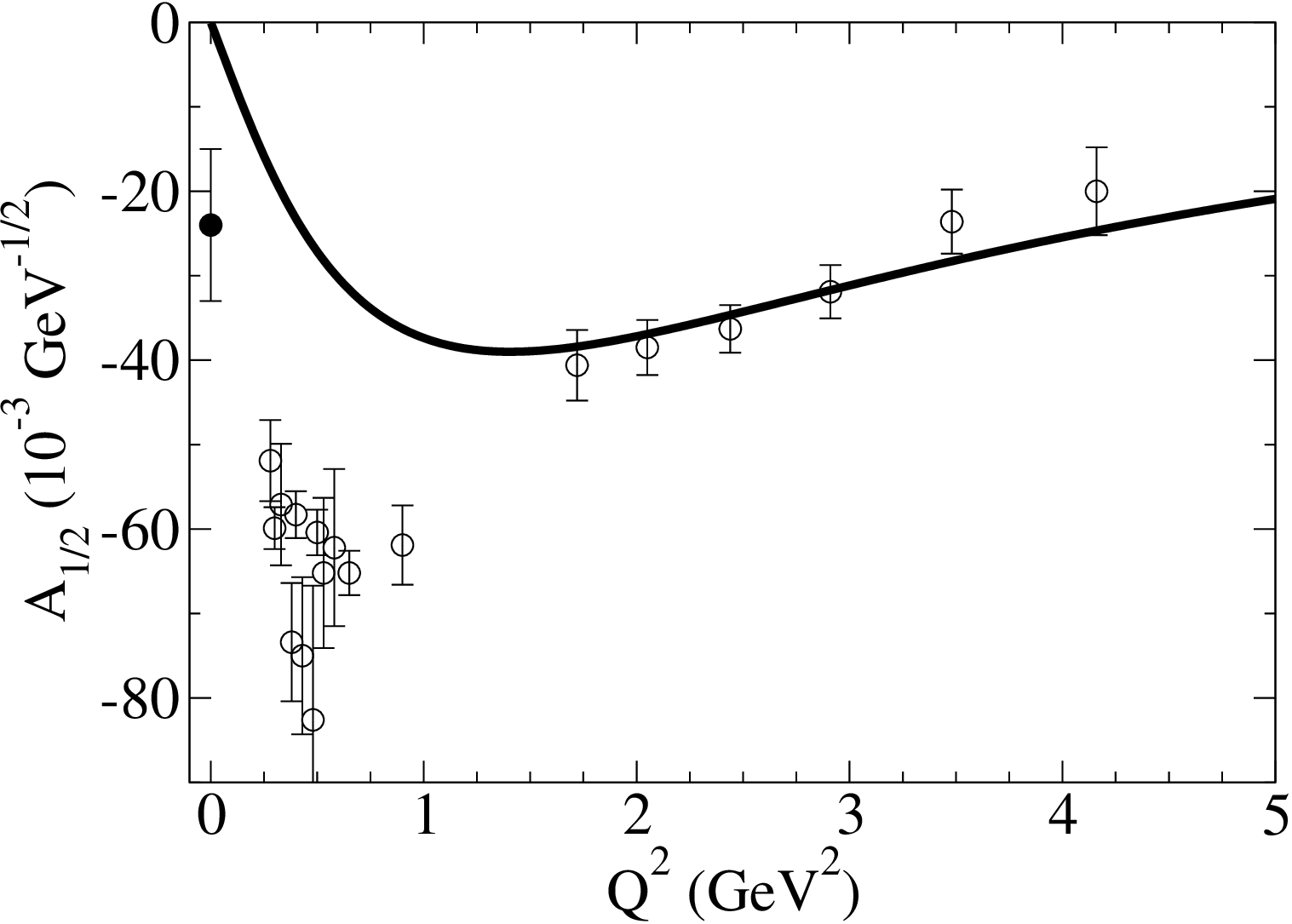}
}}
\caption{\footnotesize{
Results for the 
$\gamma^\ast p \to D_{13}^+(1520)$ amplitudes $A_{1/2}$ and $A_{3/2}$.
Data from CLAS~\cite{Aznauryan09,Mokeev12} and
PDG~\cite{PDG}. 
}}
\label{figParam2}
\end{figure}

\begin{figure}[t]
\vspace{.3cm}
%\centerline{
%\mbox{\includegraphics[width=2.8in]{A12_N1535a} }}
%\vspace{1.5cm}
\centerline{
\mbox{
\includegraphics[width=2.8in]{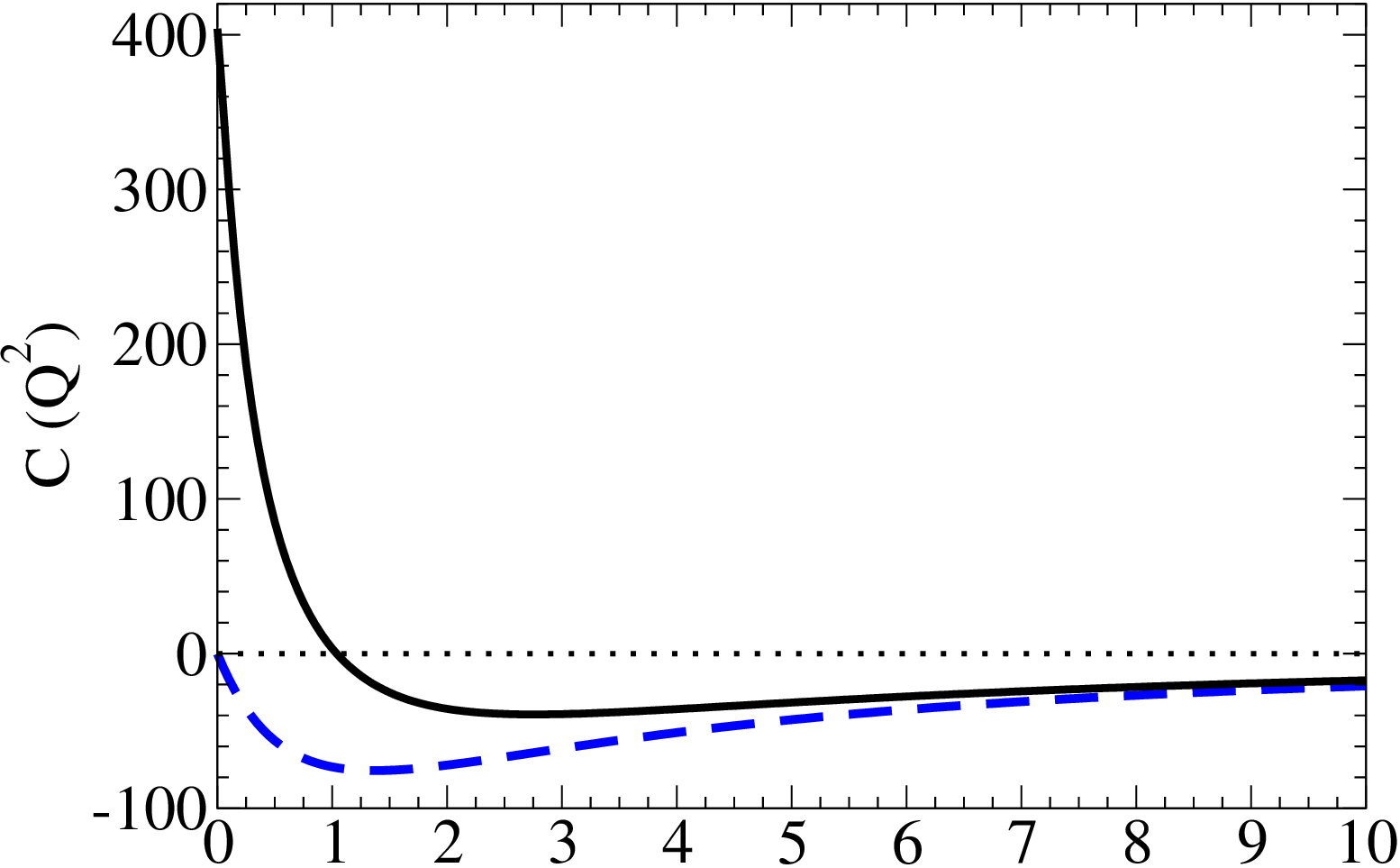}
}}
\vspace{.8cm}
\centerline{
\mbox{
\includegraphics[width=2.8in]{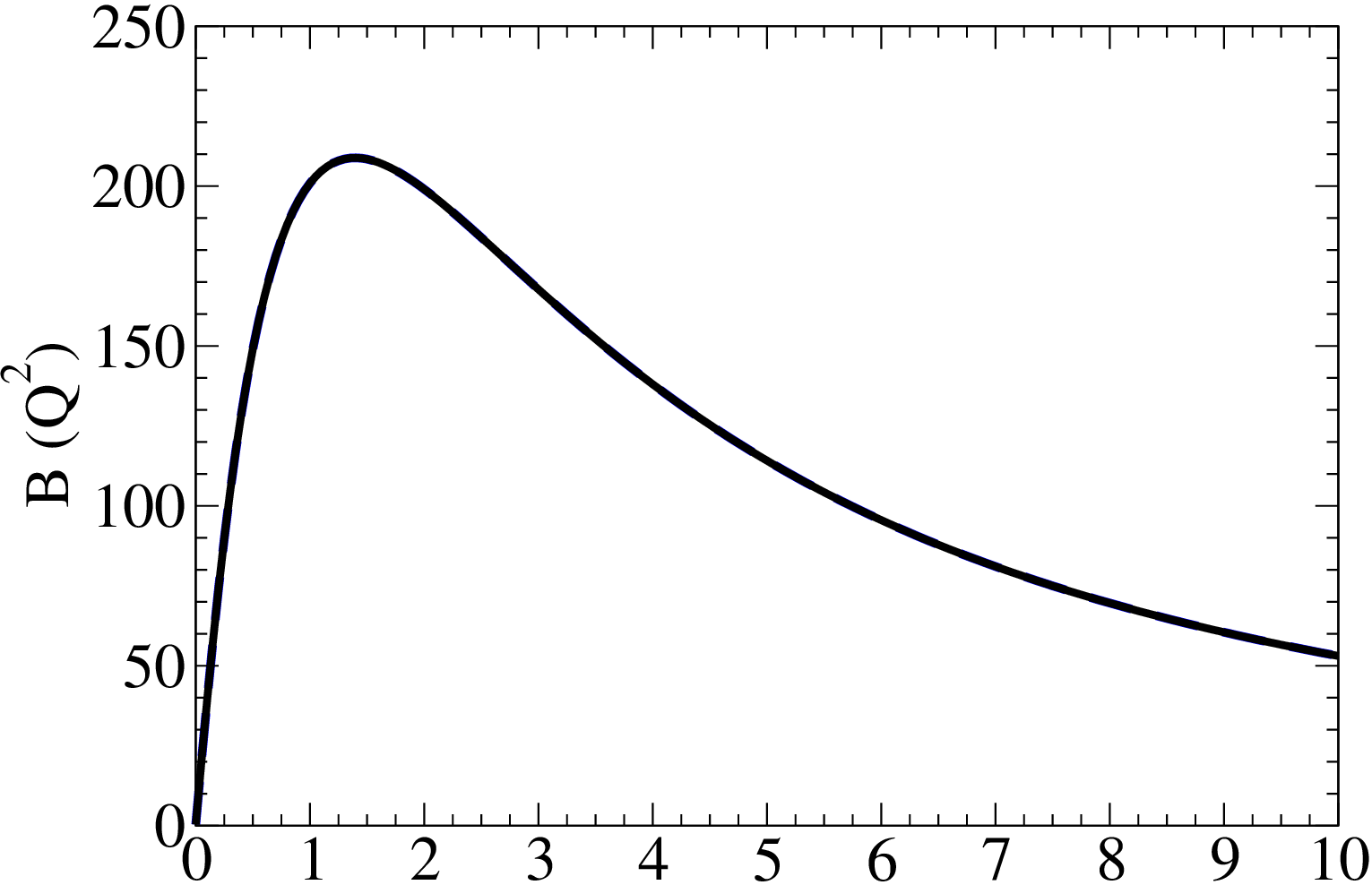}
}}
\vspace{.9cm}
\centerline{
\mbox{
\includegraphics[width=2.8in]{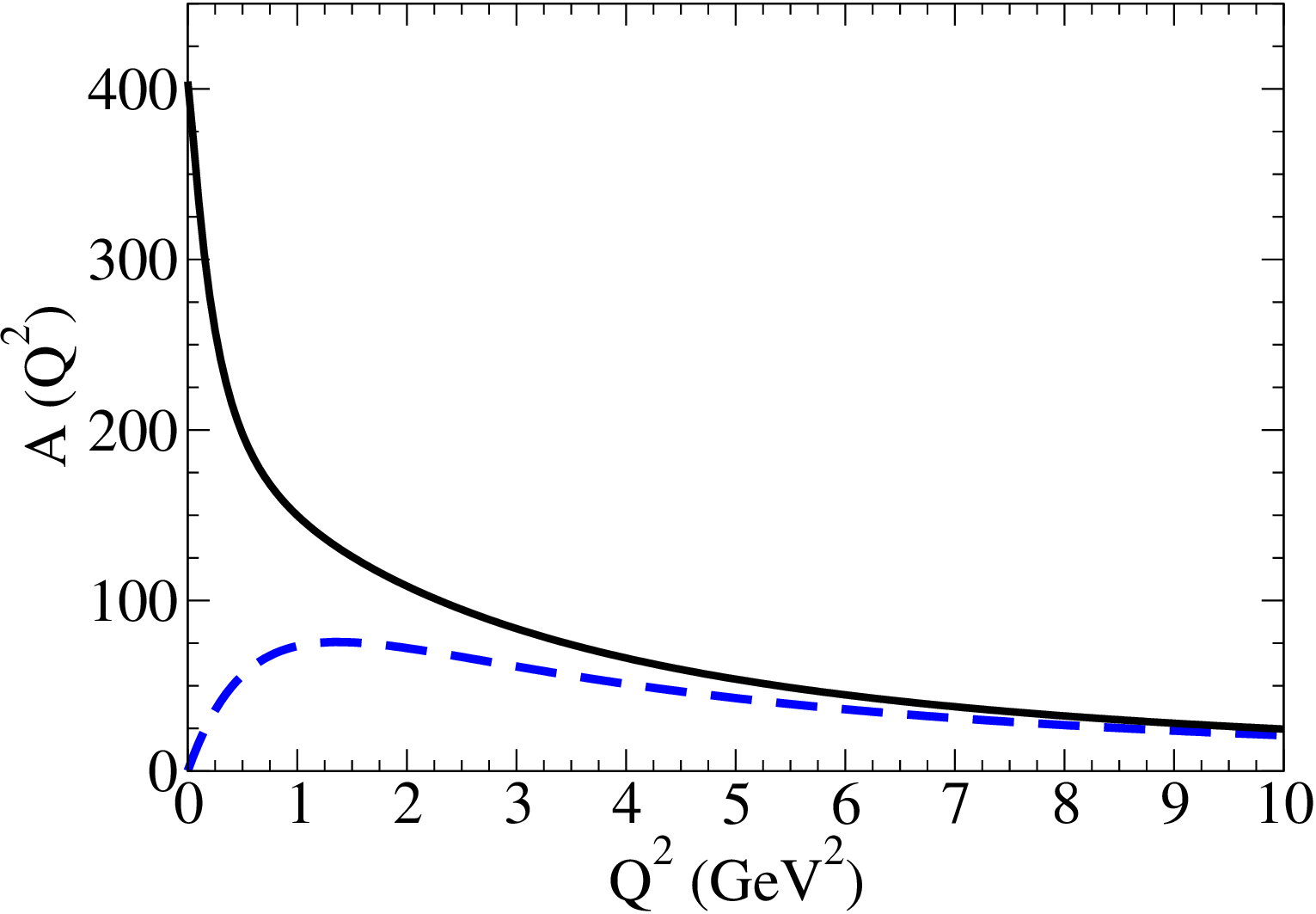}
}}
\caption{\footnotesize{
Coefficients $A,B$, and $C$ for the  model 1 (dashed line),
model 2 (solid line), 
according to Eqs.~(\ref{eqA})-(\ref{eqC}). 
For model 1, $C=-A$.
$B$ is the same in both models.}}
\label{figCoeff}
\end{figure}

\section{Results}
\label{secResults}

We will present the results in the following way:
First we show the results from
the covariant spectator quark model for the 
$\gamma^\ast N \to S_{11}(1535)$ and 
 $\gamma^\ast N \to D_{13}(1520)$ amplitudes.
Next we calculate the functions $A,B$, and $C$
using the expressions derived in Sec.~\ref{secSQTM}.
Finally we use the functions $A,B$, and $C$ 
to estimate the amplitudes $A_{1/2}$ and $A_{3/2}$ for
the remaining electromagnetic transitions.

\subsection{Model for the $\gamma^\ast N \to S_{11}(1535)$ and 
 $\gamma^\ast N \to D_{13}(1520)$ amplitudes}

The results of 
amplitude $A_{1/2}$ for the $\gamma^\ast N \to S_{11}(1535)$ 
transition, given by Eq.~(\ref{eqA12S}),  
are presented in Fig.~\ref{figParam1}.
The deviation between model and data 
at low $Q^2$ can be the consequence 
of the meson cloud not included in the model.
For the $S_{1/2}$
there are evidences that  it  is correlated 
with $A_{1/2}$ at large $Q^2$ \cite{S11scaling}.

The results of the amplitudes 
$A_{1/2}$ and $A_{3/2}$ relative for transition 
$\gamma^\ast N \to D_{13}(1520)$ are presented 
in Fig.~\ref{figParam2}. 
For the amplitude $A_{1/2}$, given by Eq.~(\ref{eqA12}) we consider 
only the valence quark contributions 
given by Eq.~(\ref{eqGM}) and drop the term $G_4^\pi$.
As for $A_{3/2}$, since the valence quark contributions 
vanishes (in the limit $\sin \theta_D \to 0$),
we present the result obtained by the parametrization 
of the meson cloud given by Eq.~(\ref{eqA32}).

From Figs.~\ref{figParam1} and \ref{figParam2}
we may conclude that one has a good description 
of the data (about two standard deviations) 
for the $\gamma^\ast N \to N^\ast(1535)$ transition when $Q^2 > 2.5$ GeV$^2$
and for the $\gamma^\ast N \to N^\ast(1520)$ transition
when $Q^2 > 1.5$ GeV$^2$.
We may then say that the covariant spectator 
quark model is reliable for $Q^2 > 2$ GeV$^2$.

\subsection{Calculation of $A,B$, and $C$}

In the calculation of the coefficients $A,B$, and $C$
we consider two different approximations:
\begin{itemize}
\item
{\bf Model 1:} 
we use the approximation 
$A_{3/2}^{D13} \equiv 0$, based on Eqs.~(\ref{eqA12S1})-(\ref{eqA12D1}),
and calculate the two independent functions ($A$ and $B$).
The results are represented by a dashed line.
In this case only valence quark degrees of freedom are considered. 
\item
{\bf Model 2:} 
we include the parametrization of the 
meson cloud effects for the amplitude $A_{3/2}^{D13}$ 
given by Eq.~(\ref{eqG4}), 
and calculate the three coefficients 
using Eqs.~(\ref{eqA})-(\ref{eqC}).
The results are represented by a solid line.
\end{itemize}

The results for coefficients $A,B$, and $C$
are presented in Fig.~\ref{figCoeff}
for model 1 (dashed line) and 
model 2 (solid line).

Note that the function $B$ is the same for both models,
since it is independent of  $A_{3/2}^{D13}$
[see Eq.~(\ref{eqB})].
For future discussion we note 
also that the difference $A-C$ is also
the same in both models 
[see Eqs.~(\ref{eqA}) and (\ref{eqC})].
From Table~\ref{tableAmp} 
we may then conclude that 
the amplitudes $A_{1/2}$ will 
be the same for both models 
for the cases $S_{11}(1650)$ and $D_{13}(1700)$.

\begin{figure}[t]
\vspace{.3cm}
%\centerline{
%\mbox{\includegraphics[width=2.8in]{A12_N1535a} }}
%\vspace{1.5cm}
\centerline{
\mbox{
\includegraphics[width=2.8in]{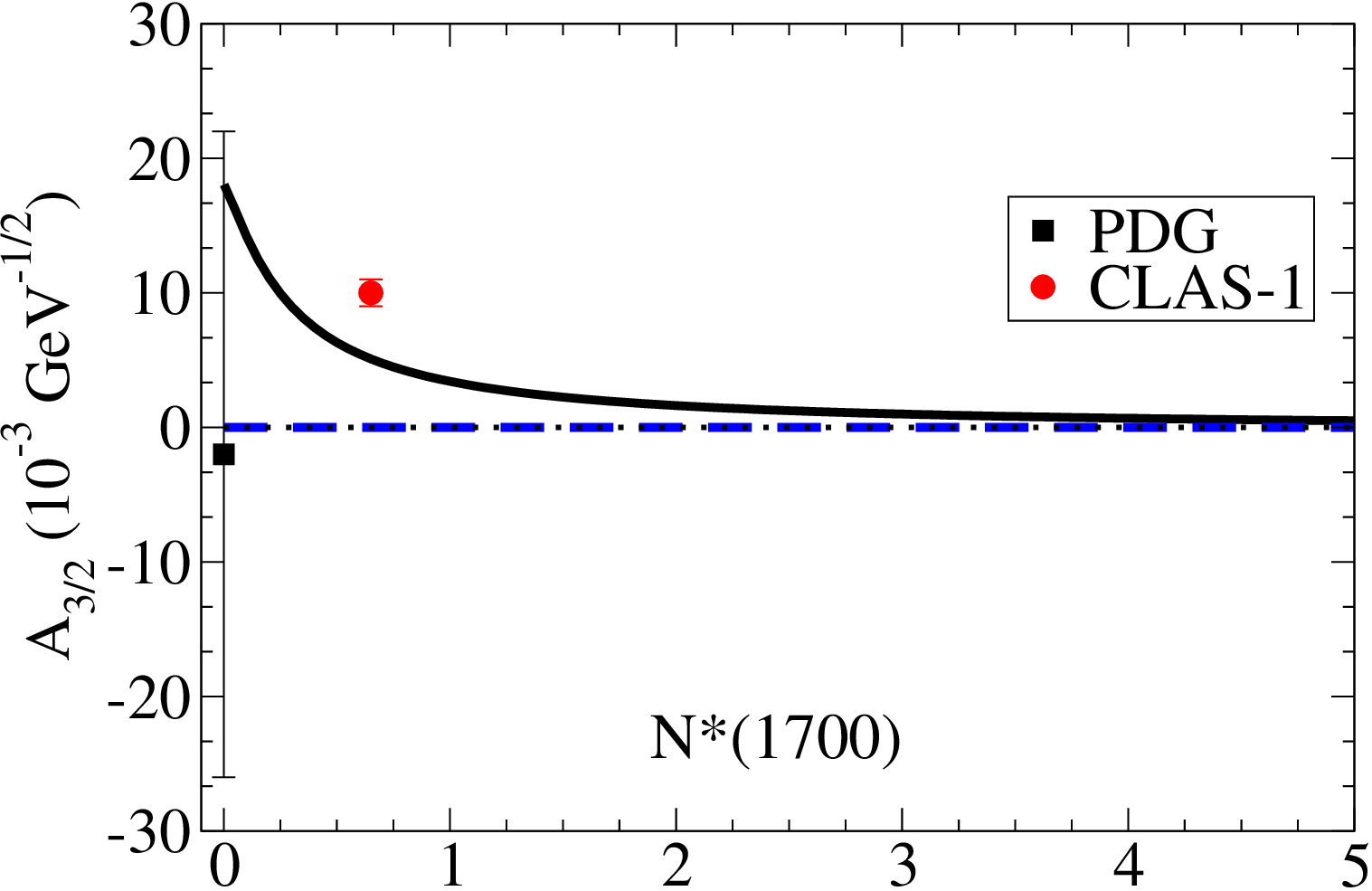}
}}
\vspace{.8cm}
\centerline{
\mbox{
\includegraphics[width=2.8in]{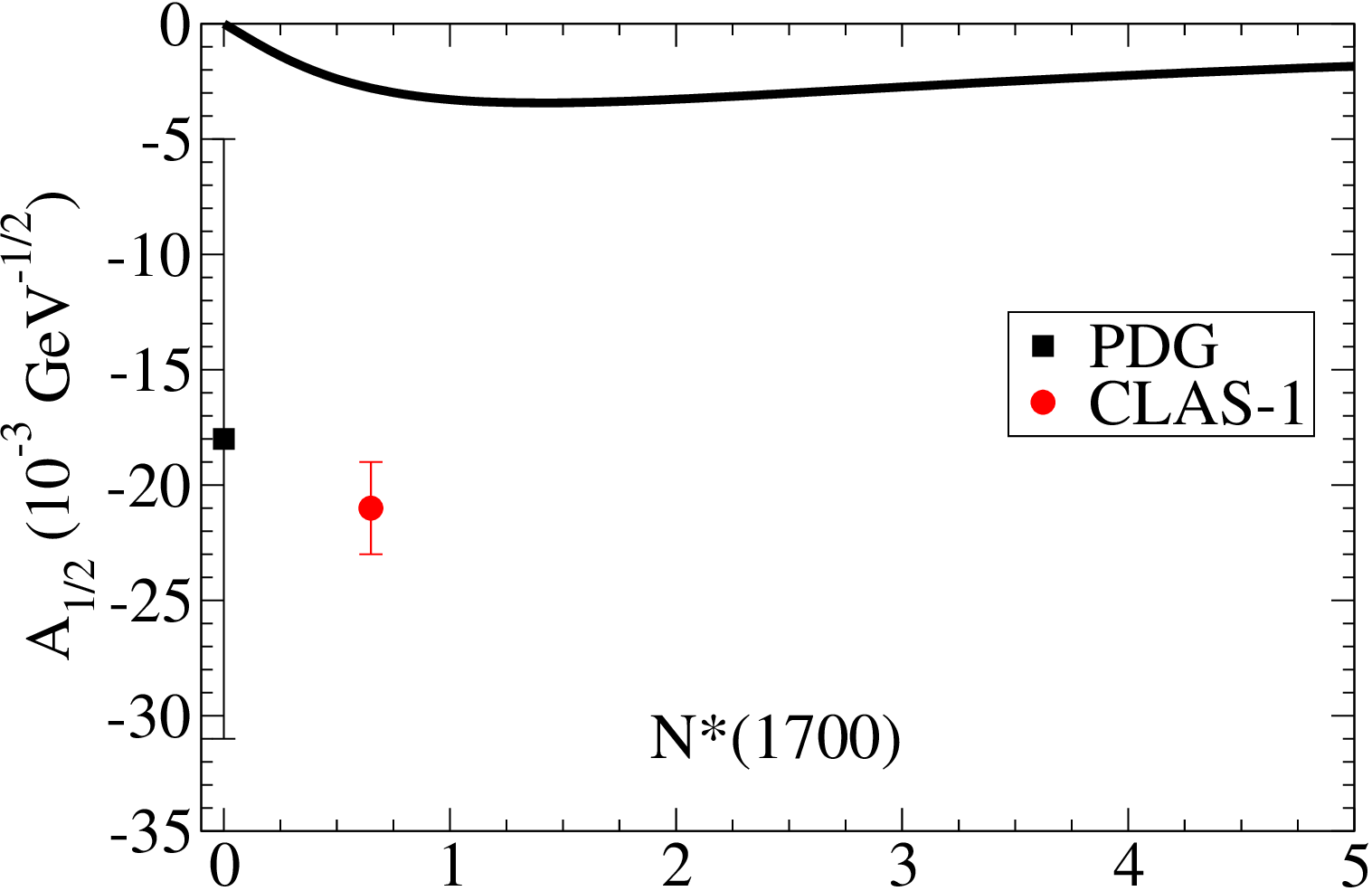}
}}
\vspace{.8cm}
\centerline{
\mbox{
\includegraphics[width=2.8in]{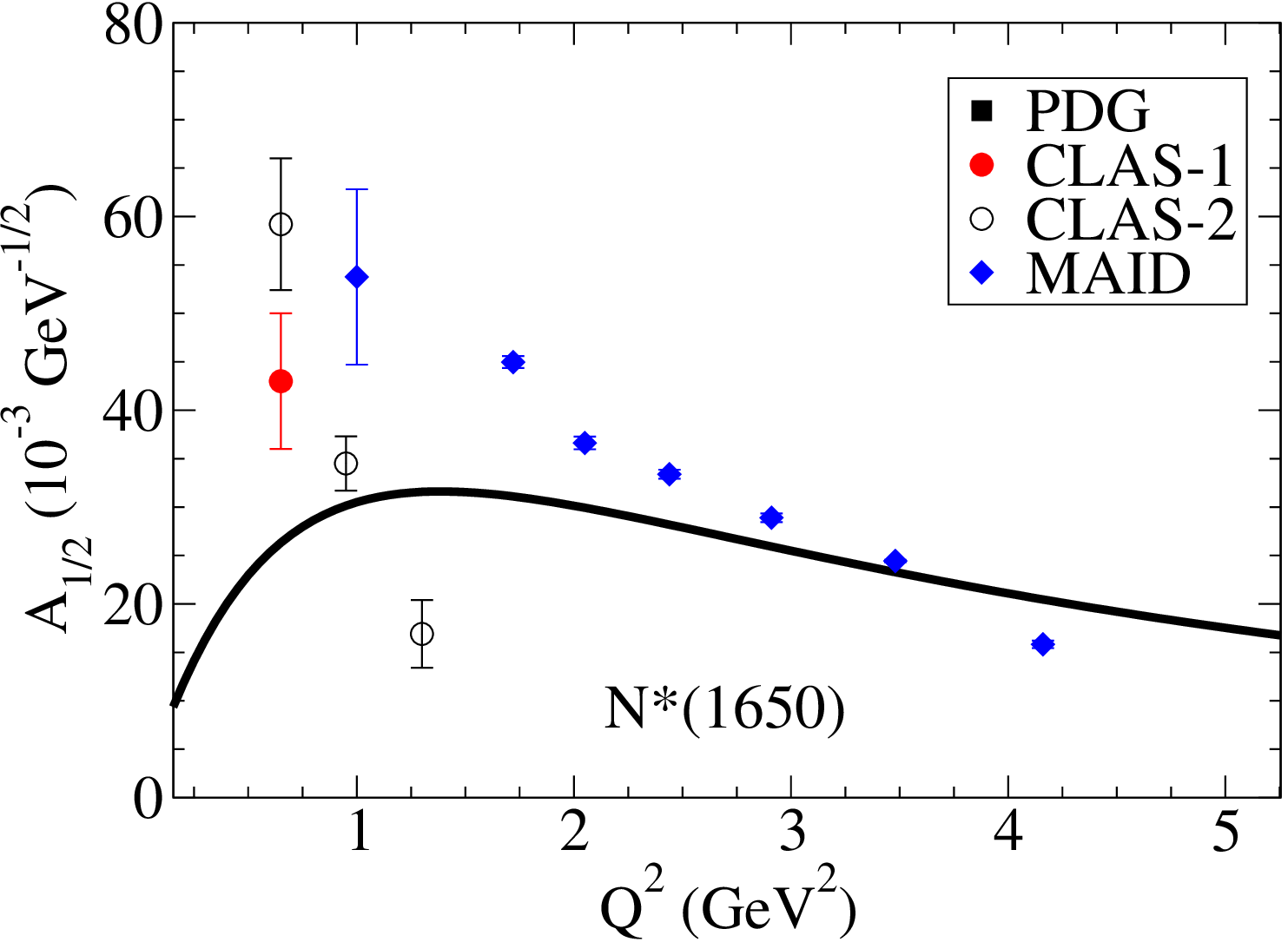}
}}
\caption{\footnotesize{
Amplitudes for the 
$\gamma^\ast p \to S_{11}^+(1650)$
and $\gamma^\ast p \to D_{13}^+(1700)$ transitions.
Model 1 (dashed line) and model 2 (solid line).
CLAS data from Refs.~\cite{Aznauryan05,Dugger09},
MAID data from Refs.~\cite{MAID1,MAID2} and 
PDG data from Ref.~\cite{PDG}. 
}}
\label{figNstar}
\end{figure}

\begin{figure}[t]
\vspace{.3cm}
\centerline{
\mbox{
\includegraphics[width=2.8in]{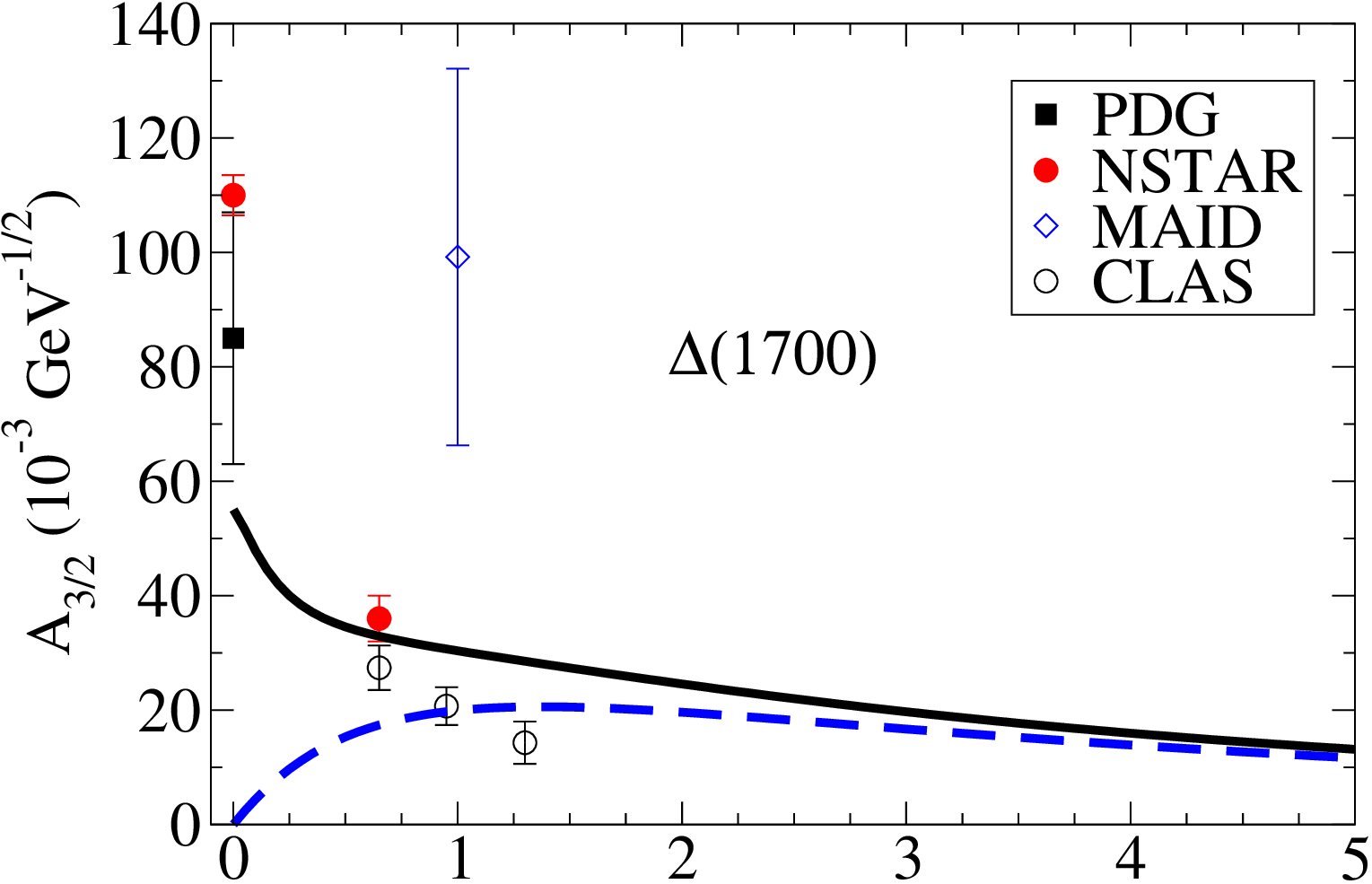}
}}
\vspace{.8cm}
\centerline{
\mbox{
\includegraphics[width=2.8in]{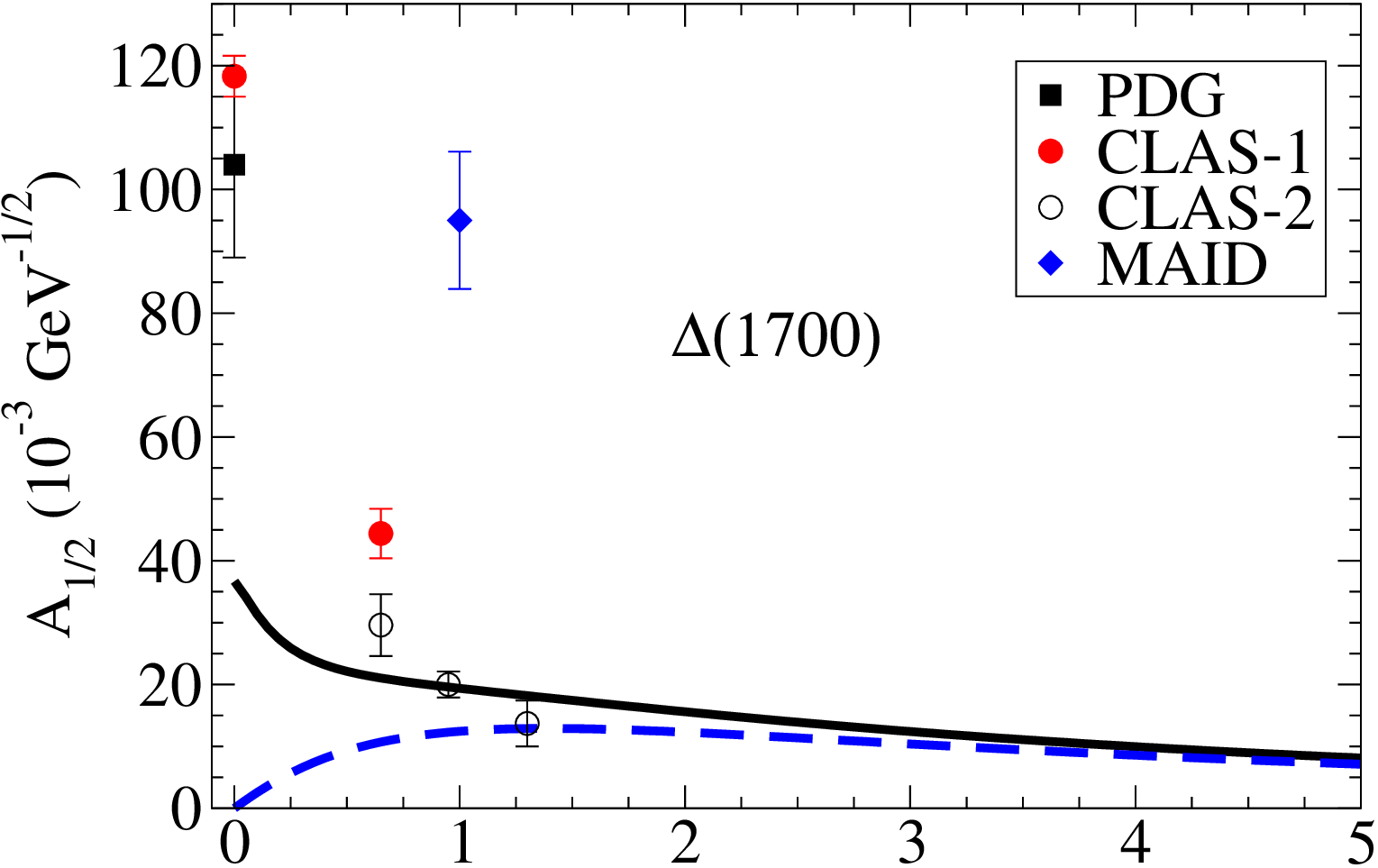}
}}
\vspace{.8cm}
\centerline{
\mbox{
\includegraphics[width=2.8in]{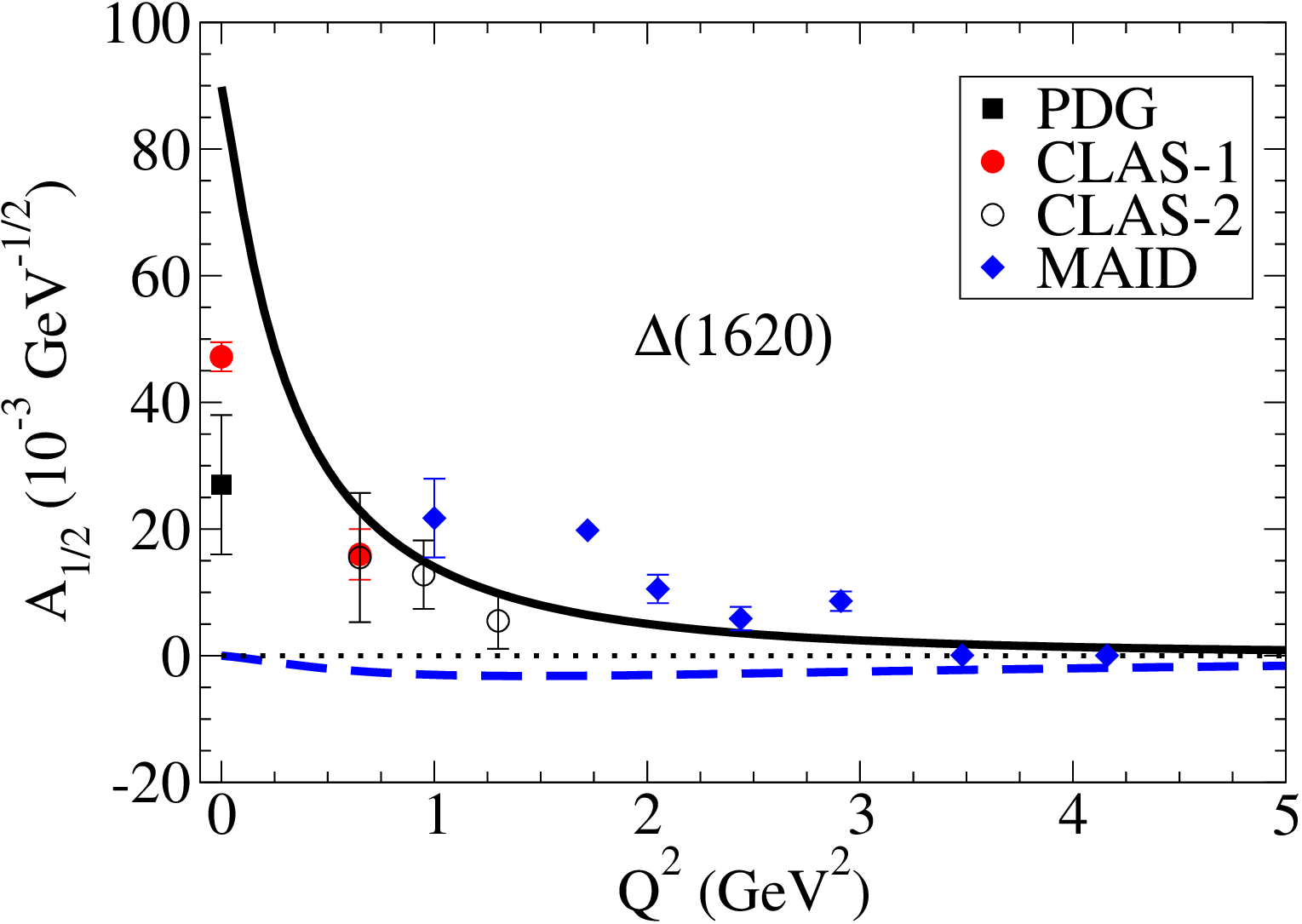}
}}
\caption{\footnotesize{
Amplitudes for the  
$\gamma^\ast p \to S_{31}^+(1620)$
and $\gamma^\ast p \to D_{33}^+(1700)$ transitions.
Model 1 (dashed line) and model 2 (solid line).
CLAS data from Refs.~\cite{Aznauryan05,Dugger09},
MAID data from Refs.~\cite{MAID1,MAID2} and 
PDG data from Ref.~\cite{PDG}. 
}}
\label{figDstar}
\end{figure}

\subsection{$\gamma^\ast N \to N^\ast$ amplitudes}

Using the parametrization from the SQTM 
of the amplitudes $A_{1/2}$ and $A_{3/2}$ 
given in Table~\ref{tableAmp}, 
and the results of the 
coefficients $A,B$, and $C$ presented 
in Fig.~\ref{figCoeff}, we can 
calculate the amplitudes for the transitions
$\gamma^\ast N \to N^\ast(1650)$,  
$\gamma^\ast N \to N^\ast(1700)$,
$\gamma^\ast N \to \Delta(1620)$,
and $\gamma^\ast N \to \Delta(1700)$,
relative to the reactions with 
proton targets.
We recall that the range of application
of the model is $Q^2 > 2$ GeV$^2$.

We compare our results with the CLAS data 
from Refs.~\cite{Aznauryan05,Dugger09},
labeled as CLAS-1; preliminary data  from CLAS, labeled as CLAS-2 
\cite{NSTAR,Mokeev_CLAS}; 
MAID data~\cite{MAID1,MAID2},
and PDG data for $Q^2=0$~\cite{PDG}.
The CLAS-1 data include data at 
the photon point~\cite{Dugger09} (single pion production)
and at $Q^2=0.65$ GeV$^2$ (double pion production)~\cite{Aznauryan05}.
Not included are the data from 
Refs.~\cite{Burkert03,NSTAR_data}
composed by single pion production only
and results presented  in proceedings,
conferences, and workshops.

The estimates based on the SQTM formalism 
for the helicity amplitudes relative to the 
$\gamma^\ast N \to S_{11}(1650)$
and $\gamma^\ast N \to D_{13}(1700)$ transitions
are presented in Fig.~\ref{figNstar}.
In the figure, models 1 and 2 
for the amplitudes $A_{1/2}$ are 
indistinguishable because $A-C$ 
is the same for both models, as discussed previously.
The amplitude $A_{3/2}$ for the $\gamma^\ast N \to D_{13}(1700)$ transition,
vanishes in model 1, because $A+C=0$
in that case (by construction). 
We may conclude then that model 1
is insufficient to describe the data.
For that reason and also 
because of the results for the  
$\gamma^\ast N \to D_{13}(1520)$ transition 
in the following we favor model 2.

From the graph 
for the $\gamma^\ast N \to S_{11}(1650)$ transition
we conclude that 
both models have the same magnitude as the data, 
although the MAID data have very small errorbars.
As for the $\gamma^\ast N \to D_{13}(1700)$ transition, 
one cannot draw too many conclusions,
since there are no data available for large $Q^2$,
except that our estimate for model 2
is close to the datapoint from CLAS-1 for $A_{3/2}$,
and it is possible that 
 model 2 can provide 
a good estimate for larger $Q^2$.
For the amplitude $A_{1/2}$ the model 
underestimate in absolute value the data 
at low $Q^2$.

The results for the  
$\gamma^\ast N \to S_{31}(1620)$
and $\gamma^\ast N \to D_{33}(1700)$ transitions
are presented in Fig.~\ref{figDstar}.
For the $\gamma^\ast N \to S_{31}(1620)$
we conclude that the model 1 
gives the wrong sign and magnitude 
for the $A_{1/2}$ amplitude.
As for the $\gamma^\ast N \to D_{33}(1700)$,
model 1 approaches the results from model 2 
when $Q^2$ increases.
Back to the $\gamma^\ast N \to S_{31}(1620)$
transition we conclude  
that model 2 is close to the data for $Q^2>$ 1 GeV$^2$
and underestimates the MAID data by about 
1-2 standard deviations,
since the errorbars are very small.
For the $\gamma^\ast N \to D_{33}(1700)$ transition,
we cannot conclude much, 
because there are no data for $Q^2> 1.5$ GeV$^2$,
except that both models
are close to the CLAS-2 data for $Q^2> 1$ GeV$^2$, 
and therefore they may be used to make projections 
for higher $Q^2$.

In order to check the predictions shown in Figs.~\ref{figNstar} and 
\ref{figDstar}, 
new data on the resonances with masses above 1.6 GeV are needed 
for $Q^2> 2$ GeV$^2$, where the estimate for quark core contributions 
can be confronted to the data. 
The data on double charged pion electroproduction 
for $Q^2=2$--5 GeV$^2$ expected from the experiments with 
the CLAS detector, will allow us for the first time 
to explore most of the resonances  
in a mass range up to 2.0 GeV 
for $Q^2< 5 $ GeV$^2$~\cite{UsersM_Mokeev}.

\subsection{Parametrization for high $Q^2$}

\begin{table}[t]
\begin{tabular}{c c c c}
\hline
\hline
State & Amplitude &  $D(10^{-3}$GeV$^{-1/2}$) & $\Lambda^2$(GeV$^2$)   \\
\hline
$S_{11}(1650)$ & $A_{1/2}$ & 68.90 &  3.35\\ [.3cm]
$S_{31}(1620)$ & $A_{1/2}$ &  ... & \sp ... \\ [.3cm]
$D_{13}(1700)$ & $A_{1/2}$ &  $-8.51$\sp\sp & 2.82 \\
& $A_{3/2}$ & 4.36 & 3.61   \\ [.3cm]
$D_{33}(1700)$ & $A_{1/2}$ &  39.22  & 2.69 \\ 
& $A_{3/2}$ & 42.15 & 8.42   \\ [.1cm]
\hline
\hline
\end{tabular}
\caption{
Parameters from the high $Q^2$ parametrization,
according to Eqs.~(\ref{eqA12LQ2})-(\ref{eqA32LQ2}).}
\label{tableLargeQ2}
\end{table}

In order to facilitate the comparison with 
future experimental data, we derived 
simple analytic approximations 
for our numerical results at large $Q^2$.
Based in the expected large $Q^2$ behavior: 
$A_{1/2} \propto 1/Q^3$ and 
$A_{3/2} \propto 1/Q^5$ \cite{Carlson},
we choose for large $Q^2$,
the forms
\ba
& &
A_{1/2}(Q^2) = D \left( \frac{\Lambda^2}{
\Lambda^2 + Q^2}\right)^{3/2} 
\label{eqA12LQ2}\\
& &
A_{3/2}(Q^2) = D \left( \frac{\Lambda^2}{
\Lambda^2 + Q^2}\right)^{5/2}.
\label{eqA32LQ2}
\ea
In the previous expressions $D$ is a 
coefficient and $\Lambda$ a cutoff
that depend on the amplitude 
($A_{1/2}$ or $A_{3/2}$) and transition.
We note however that these parametrizations 
are valid for a restricted 
region of $Q^2$, since in the covariant 
spectator quark model the form factors 
and amplitudes are affected by logarithm corrections
in $Q^2$ for 
large $Q^2$~\cite{NDelta,S11,D13}. 

All amplitudes, except for the $\gamma^\ast N \to S_{31}(1620)$ transition,
are well described by the analytic forms 
of Eqs.~(\ref{eqA12LQ2})-(\ref{eqA32LQ2}).
The numerical results for $D$ and $\Lambda^2$ are presented in 
Table \ref{tableLargeQ2}.
The parameters were calculated 
in order to reproduce the results 
from model 2 exactly for $Q^2=5$ GeV$^2$,
but they also provide good approximations for 
values of $Q^2$ up to 10 GeV$^2$, or even larger.

We found out that the amplitude $A_{1/2}$ for the 
transition  $\gamma^\ast N \to S_{31}(1620)$ 
cannot be approximated by Eq.~(\ref{eqA12LQ2}),
in particular by the power $3/2$.
The falloff of that amplitude is consistent 
with a stronger falloff.
In order to interpret that result 
we start noting that we can write,
using Table~\ref{tableAmp} and Eqs.~(\ref{eqA})-(\ref{eqC}),
\ba
A_{1/2}^{S31} \propto 
\left(2 \frac{A_{1/2}^{S11}}{\cos \theta_S} 
+ 4 \sqrt{2} A_{1/2}^{D13} + 4 \sqrt{6} A_{3/2}^{D13}
\right).
\label{eqA12_S31}
\ea
If the amplitudes $A_{1/2}^{S11}, A_{1/2}^{D13} \propto 1/Q^3$,
as expected, the deviation from  $A_{1/2}^{S31}$
from $1/Q^3$ is a consequence of a partial 
cancellation between the two amplitudes $A_{1/2}$ 
in Eq.~(\ref{eqA12_S31}).
This interpretation makes sense
because those amplitudes 
have different signs (see Figs.~\ref{figParam1} and \ref{figParam2}). 
Due to the cancellation between the leading order term ${\cal O}(1/Q^3)$ 
of the first two terms, 
$A_{1/2}^{S31}$  is dominated 
by the second order terms and the amplitude $A_{3/2}^{D13}$.
A simple empirical parametrization
of the amplitude in units of
$10^{-3}$ GeV$^{-1/2}$  
is $A_{1/2}^{S31}= 77.21 \left ( \frac{\Lambda^2}{\Lambda^2+ Q^2}\right)^{5/2}$,
with $\Lambda^2= 1$ GeV$^2$.

\section{Summary and conclusions}
\label{secConclusions}

In this work we combined the features 
of the covariant spectator quark model 
and the single quark transition model 
to predict the transition 
amplitudes $A_{1/2}$, $A_{3/2}$ for 
the transitions 
$\gamma^\ast N \to S_{11}(1650)$, 
$\gamma^\ast N \to D_{13}(1700)$,
$\gamma^\ast N \to S_{31}(1620)$, and 
$\gamma^\ast N \to D_{33}(1700)$.
The resonances in the 
final state are all 
members of the $[70,1^-]$ supermultiplet.
We follow the method used 
in Refs.~\cite{Aznauryan12,Burkert03,Burkert04},
but we use a theoretical model (quark model) 
to extract the characteristic coefficients 
associated with the transition,
instead of the experimental data,
which are contaminated by meson cloud 
effects at small $Q^2$.

Since the covariant spectator quark model 
and the SQTM are based 
in the valence quark degrees of freedom, 
the range of application of the models
is the region of intermediate and high $Q^2$.
In the present case 
we may define that region as $Q^2>2$ GeV$^2$,
based on the results for the transitions 
used in the calibration of the SQTM model.
The region $Q^2>2$ GeV$^2$ 
is where meson cloud effects 
are expected to play a minor role.
However, since the covariant spectator quark model 
predicts that $A_{3/2} =0$ 
for the  $\gamma^\ast N \to D_{13}(1520)$ transition,
we explore the possibility of improving our estimations
including a meson cloud parametrization of that amplitude.
We recall that the proposed parametrization for the  
meson cloud contribution has a very slow falloff 
(large cutoff $\Lambda_4^2$),
which is more typical of a valence quark contribution 
than from a meson cloud effect contribution. 
The inclusion of a parametrization 
of the amplitude $A_{3/2}$ allows 
a much better description of the data 
for intermediate $Q^2$,
although for much larger $Q^2$ 
($Q^2 > 5$ GeV$^2$),
the models with  $A_{3/2} = 0$
and  $A_{3/2} \ne 0$ are very similar.

To facilitate the comparison with 
future experimental data at high $Q^2$,
we present also simple parametrizations 
of the amplitudes $A_{1/2}$ and $A_{3/2}$ 
for the different transitions, 
compatible with the expected 
falloff at high $Q^2$:
$A_{1/2} \propto 1/Q^3$, $A_{3/2} \propto 1/Q^5$.
The exception to the previous rules 
is the amplitude $A_{1/2}$ 
for the $\gamma^\ast N \to S_{31}(1620)$ transition,
where we predict a falloff faster than $1/Q^3$.

Summarizing,  we present predictions for the $[70,1^-]$ 
amplitudes in the region  $Q^2> 2$  GeV$^2$, 
where we can expect a dominance of 
the valence quark degrees of freedom.
Nevertheless the  meson cloud contributions 
may still be important for some electromagnetic transitions.
In addition we present parametrizations 
for the region $Q^2 \approx 5$ GeV$^2$, or larger,
that can be tested in the future JLab-12 GeV upgrade.

\begin{acknowledgments}
The author wants to thank Viktor Mokeev and Ralf Gothe 
for reading the manuscript and for the 
helpful discussions, comments and suggestions.
The author thanks also  Viktor Mokeev 
for sharing the preliminary data from CLAS, 
labeled here as CLAS-2.
This work was supported by the Brazilian Ministry of Science,
Technology and Innovation (MCTI-Brazil).
\end{acknowledgments}

%\newpage

\appendix

\section{New parametrization for the $S_{11}(1535)$ amplitudes}
\label{appS11}

The $S11$ system and the 
$\gamma^\ast N \to S_{11}(1535)$ transition were studied
in Ref.~\cite{S11} within the 
covariant spectator quark model formalism.
In that paper it was assumed that 
the diquark was pointlike.
If we use a more detailed treatment,
where nonpointlike diquark states are considered, 
following the formalism of Ref.~\cite{D13}, 
one has to  correct the normalization factor from $N=\frac{1}{2}$ to 
$N=\frac{1}{\sqrt{2}}$, reducing the first estimate 
by a factor $1/\sqrt{2}$.
The final expression 
for the Dirac form factor $F_1^\ast$ is then
\ba 
F_1^\ast = \frac{\sqrt{2}}{3} 
\left( f_{1+} + 2 f_{1-} \tau_3 \right) {\cal I}_{S11}.
\label{eqF1}
\ea

Another aspect that can be improved in 
the model from Ref.~\cite{S11} is the low $Q^2$ 
region dependence of the form factors.
In the model from  Ref.~\cite{S11},
the nucleon and the $S11$ states were not strictly orthogonal
therefore the model failed near $Q^2=0$,
because ${\cal I}_{S11}(0) \ne 0$ and $A_{1/2} (0) \to \infty$. 
The exact  orthogonality between those states 
demands ${\cal I}_{S11}(0) = 0$.
We can fix that problem redefining the 
radial wave function in order to 
obtain ${\cal I}_{S11}(0) = 0$.
The price to pay is the introduction 
of a new parameter in the radial wave function
$\psi_{S11}$, which can be fixed by 
a fit to the data, as explained next.
The same procedure was used in Ref.~\cite{D13}
for the $D_{13}(1520)$ wave function.

We note however that even in the present case, 
where the model is valid near $Q^2=0$, 
we cannot expect a very good agreement 
between model and experimental data 
at low $Q^2$, because the meson cloud 
effects are not included.

\subsection{Imposing the orthogonality between states}

In general, in the covariant spectator quark model
the radial wave functions can be expressed 
in terms of the variable $(P-k)^2$,
where $P$ and $k$ are respectively the 
baryon and the diquark momenta,
because the baryon and the diquark 
are both on-mass-shell. 
The dependence in the momenta
can then be represented using 
the dimensionless variable
\ba
\chi= \frac{(M_B -m_D)^2 -(P-k)^2}{M_B m_D},
\ea 
where $M_B$ and $m_D$ are respectively 
the baryon and diquark masses.

In that case
the nucleon wave function is defined as
\ba
\psi_N(P,k)= \frac{N_0}{m_D ( \beta_1 + \chi)(\beta_2 + \chi)},
\label{eqPsiN}
\ea
where $N_0$ is a normalization constant and 
$\beta_1,\beta_2$ are parameters  
determined in Ref.~\cite{Nucleon} by a fit 
to the nucleon form factor data (model II).
The numerical values are 
$\beta_1= 0.049$ and $\beta_2 = 0.717$.
As $\beta_2 > \beta_1$, $\beta_1$ 
parametrizes the long range region and 
 $\beta_2$ the short range region, 
in the coordinate space.

The overlap integral (\ref{eqIS11}) includes 
also the $S11$ radial wave function.
In Ref.~\cite{S11} we define $\psi_{S11}$ by 
the same expression given for the nucleon
by Eq.~(\ref{eqPsiN}), except for the momentum and mass
of the baryon.
The problem of that choice is that 
the integral ${\cal I}_{S11}(0)$ does not vanish, except in the case 
$M_S=M$ (nucleon and $S11$ with the same mass).
The reason why ${\cal I}_{S11}(0) \ne 0$, unless $M_S=M$, 
is because $S11$ and the nucleon cannot be at rest in the same frame.
See Ref.~\cite{S11} for a complete discussion.

We can avoid that problem, defining $\psi_{S11}$ 
in order to be orthogonal to the nucleon.
We consider then the form
\ba
\psi_{S11}(P,k)=
\frac{N_1}{m_D(\beta_2 + \chi)}
\left[
\frac{1}{\beta_1 + \chi} - \frac{\lambda_{S11}}{\beta_3 + \chi}
\right],
\ea
where $\beta_3$ is a new parameter and 
$N_1$ is a new  normalization constant.
$\lambda_{S11}$ is a parameter that 
can be fixed, once $\beta_3$ is known by
the orthogonality condition
\ba
{\cal I}_{S11}(0)=0.
\ea
The free parameter $\beta_3$ can be determined 
by the fit to the high $Q^2$ data 
from the  $\gamma^\ast N \to S_{11}(1535)$ transition.

The normalization conditions are
\ba
\int_k |\psi_N(\bar P,k)|^2=1, \hspace{.5cm} 
\int_k |\psi_{S11}(\bar P,k)|^2=1, 
\ea
where $\bar P$ represents in each case 
the baryon momentum at the baryon rest frame.

\subsection{Fit to the data}

\begin{figure}[b]
\vspace{.3cm}
\centerline{
\mbox{
\includegraphics[width=2.8in]{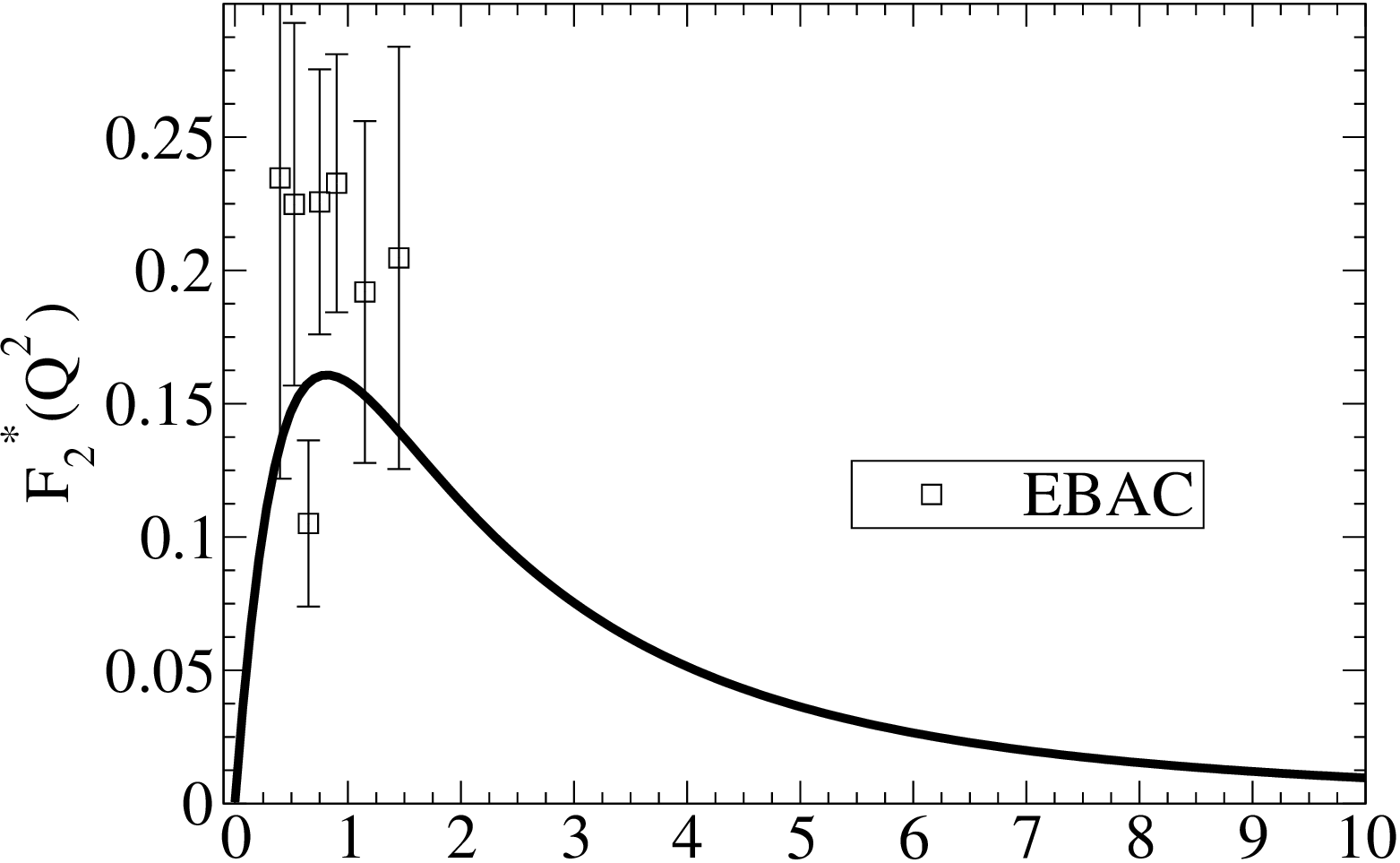}
}}
\centerline{
\vspace{.5cm} }
\centerline{
\mbox{
\includegraphics[width=2.8in]{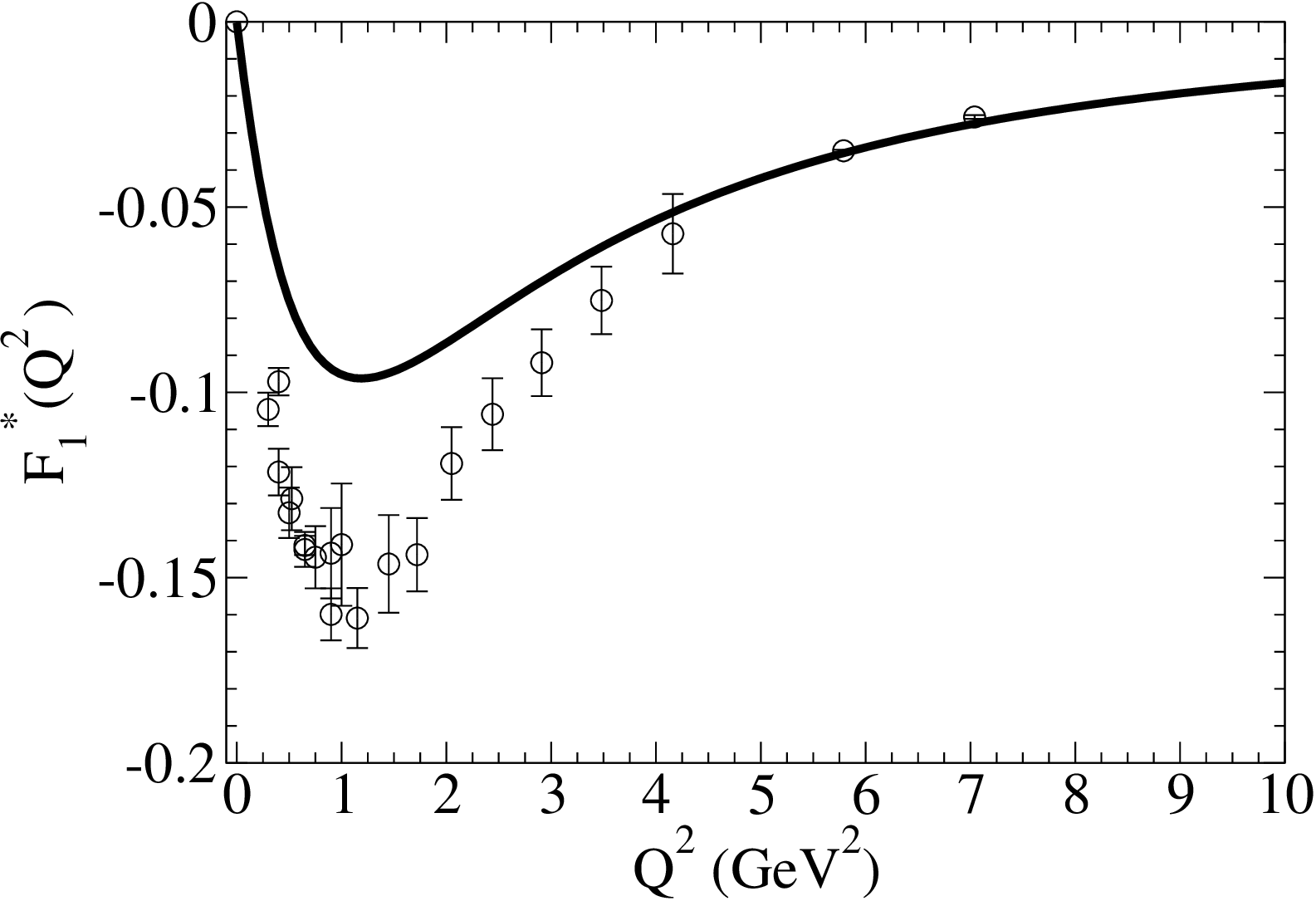}
}}
\caption{\footnotesize{
$\gamma^\ast N \to N^\ast(1535)$ form factors.
Data from CLAS~\cite{Aznauryan09} and 
JLab/Hall C~\cite{Dalton09}.
The EBAC calculations are from Ref.~\cite{EBAC}.
}}
\label{figFFS11}
\end{figure}

In order to extend the application 
of the model for the $S_{11}(1535)$ state, 
also to the low $Q^2$ regime,
we fit Eq.~(\ref{eqF1}) to the form factor data.
The only adjustable parameter available 
is the value of $\beta_3$ in the $\psi_{S11}$ 
radial wave function,
included in the integral ${\cal I}_{S11}$. 
As we are taking into account only the valence quark 
degrees of freedom, we cannot expect a good agreement 
for small $Q^2$, therefore we fit only the high $Q^2$ data.
We consider therefore only the data with $Q^2> 1.5$ GeV$^2$.

Our database includes the data from CLAS 
\cite{Aznauryan09} ($Q^2= 0.3$--$4.2$ GeV$^2$)
and from JLab/Hall C \cite{Dalton09} ($Q^2=5.4,7.0$ GeV$^2$).
The data from Hall C~\cite{Dalton09}, 
include only the amplitude $A_{1/2}$, 
assumes that $S_{1/2}$ is very small, 
and has very small errorbars.
In the fit we double the errorbars from Hall C
in order to avoid a high weight from the Hall C data.

The best fit is obtained for $\beta_3=0.540$.
The results for the form factor $F_1^\ast$ 
are presented in Fig.~\ref{figFFS11},
compared with the data mentioned previously, 
and the data from the MAID  analysis~\cite{MAID1,MAID2}.
In the same figure we show also the 
results from the valence quark contributions
for the Pauli form factor $F_2^\ast$.
We recall that in that case the experimental data 
(not shown in the graph) is consistent 
with zero for $Q^2> 1.5$ GeV$^2$.
Although one cannot compare directly 
our model for $F_2^\ast$ with the data 
due to the lack of meson cloud effects 
we can compare it with other estimates 
of the quark core effects.
In the graph for $F_2^\ast$ we present 
therefore the estimate of the bare core effects given 
by the EBAC/JLab model \cite{EBAC}.
The EBAC model is a coupled-channel reaction 
model that takes into account the 
meson and photon coupling with the baryon cores.
The result presented in the graph 
is obtained when the meson cloud effects are removed.
As we can see in the graph
our results for $F_2^\ast$ are very close to the EBAC estimates.
That result is remarkable, since no fit relative to the 
function  $F_2^\ast$ was considered.

\end{document}